\documentclass[aps,prb,twocolumn,superscriptaddress]{revtex4-2}
\usepackage[english]{babel}
\usepackage{bm}
\usepackage{amsmath}
\usepackage{graphicx}
\usepackage[colorlinks=true, allcolors=blue]{hyperref}
\usepackage{diagbox}
\usepackage{multirow}
\usepackage{threeparttable}

\begin{document}
\title{A scalable network model for electrically tunable ferroelectric domain structure in twistronic bilayers of two-dimensional semiconductors}
\author{V.~V.~Enaldiev}
\affiliation{University of Manchester, School of Physics and Astronomy, Oxford Road, Manchester M13 9PL, United Kingdom}
\affiliation{National Graphene Institute, University of Manchester, Booth St. E. Manchester M13 9PL, United Kingdom}
\affiliation{Kotelnikov Institute of Radio-engineering and Electronics of the RAS, Mokhovaya 11-7, Moscow 125009, Russia}

\author{F. Ferreira}
\affiliation{University of Manchester, School of Physics and Astronomy, Oxford Road, Manchester M13 9PL, United Kingdom}
\affiliation{National Graphene Institute, University of Manchester, Booth St. E. Manchester M13 9PL, United Kingdom}
\author{V.I. Fal'ko}
\affiliation{University of Manchester, School of Physics and Astronomy, Oxford Road, Manchester M13 9PL, United Kingdom}
\affiliation{National Graphene Institute, University of Manchester, Booth St. E. Manchester M13 9PL, United Kingdom}
\affiliation{Henry Royce Institute, University of Manchester, Booth St. E. Manchester M13 9PL, United Kingdom}

\begin{abstract}
Moir\'e structures in small-angle-twisted bilayers of two-dimensional (2D) semiconductors with a broken-symmetry interface form arrays of ferroelectric (FE) domains with periodically alternating out-of-plane polarization. Here, we propose a network theory for the tunability of such FE domain structure by applying an electric field perpendicular to the 2D crystal. Using multiscale analysis, we derive a fully parametrized string-theory-like description of the domain wall network (DWN) and show that it undergoes a qualitative change, after the arcs of partial dislocation (PD) like domain walls merge (near the network nodes) into streaks of perfect screw dislocations (PSD), which happens at a threshold displacement field dependent on the DWN period.
\end{abstract}
\maketitle

Two-dimensional material twistronics, fueled by discoveries of new phenomena in twisted graphene bilayers \cite{cao2018unconventional,cao2018correlated,Yankowitz2019,lu2019superconductors,sharpe2019emergent,cao2021nematicity,Kerelskye2021,Xu2019,Gadelha2021,Kazmierczak2021} and trilayers \cite{polshyn2020,Xu2021,Chen2019,shen2020,chen2021,park2021}, has recently expanded onto a broader range of van der Waals systems  \cite{kunstmann2018momentum,rivera2018interlayer,nayak2017probing,alexeev2019,McGilly2020,Zhang2020,Wang2020}. In general, twistronic structures are associated with geometrical moir\'e patterns: a periodic variation local stacking of the two layers. In long-period moir\'e patterns, characteristic for small-angle-twisted bilayers, the areas of energetically preferential stacking expand into mesoscale domains \cite{AldenPNAS,yoo2019atomic,rosenberger2020,Weston2020}, embedded into a domain wall network (DWN). In particular, by assembling a homobilayer of two inversion-asymmetric honeycomb monolayers (hBN or transition metal dichalcogenides (TMD)) with parallel orientation of their unit cells, one obtains a triangular array of domains with broken mirror and inversion symmetries \cite{Sung2020,woods2021,yasuda2021,stern2021,weston2021,wang2021} and an out-of-plane ferroelectric (FE) polarization \cite{li2017,WangYao2020,Ferreira2021}, which direction alternates between the neighboring XM-stacking domains (metal atoms overlaying chalcogens) and their MX twins (chalcogens over metal atoms) \cite{CarrPRB2018,Enaldiev_PRL}. 

%%%%%%%%%%%%%%%%%%%%%%%%%%%%%%%%%%%%%%%%%%%%%%
\begin{figure}%[!h]
\includegraphics[width=1.0\linewidth]{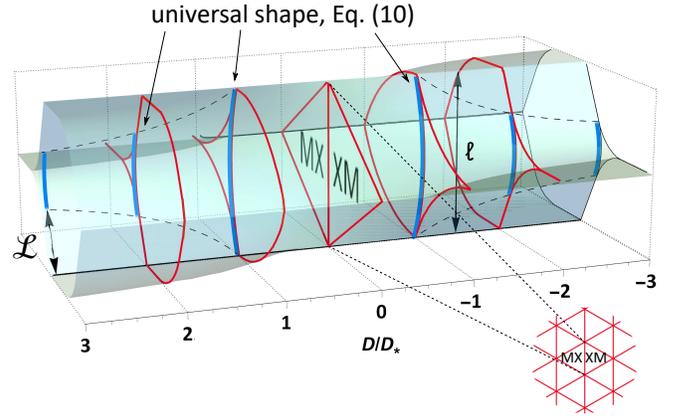}
\caption{\label{Fig0} Triangular XM/MX DWN (inset) in twisted TMD bilayers with FE domains areas varied by an out-of-plane displacement field, $D$. At $D\geq D_*$, XM/MX domain walls merge into streaks of perfect screw dislocations separating - near the network nodes - domains with the same FE polarization. The remaining arcs of XM/MX domain walls have a universal shape, scaled with $D_*/D$ ratio.  
} \end{figure}
%%%%%%%%%%%%%%%%%%%%%%%%%%%%%%%%%%%%%%%%%%%%%

The coupling between the out-of-plane FE polarization and an externally controlled displacement field, $D$, varies the energies of XM and MX, changing the ratio between their areas and leading to the deformation of DWN. Here, we offer a generic theory for the field-tunable FE domain structure in twistronic TMD bilayers, fully quantified with the help of multiscale modelling approach \cite{Enaldiev_PRL,Enaldiev_2021,Ferreira_APL,Ferreira2021} for MX$_2$-bilayers (M=Mo,W; X=S,Se). For weak fields $|D|<D_*$, the continuously deforming domain walls retain their partial dislocation character, however, above the threshold, \mbox{$|D|/D_*\geq 1$}, pairs of partial dislocations (PD) form streaks of perfect (full) screw dislocations (PSD, of length $\mathcal{L}$) near the network nodes. In Fig. \ref{Fig0}, we illustrate how the arcs of XM/MX domain walls with a universal shape merge and split apart upon the variation of $D$, where the threshold field and DWN parameters scale with elastic parameters of a partial dislocation and the DWN period, $\ell$, as $D_*\propto 1/\ell$. 

The scenario of the DWN transformation in Fig. \ref{Fig0} is a result of the following analysis. First, we use density functional theory (DFT) to quantify the coupling of FE polarization of an asymmetric TMD interface to an external out-of-plane displacement field, $D$. By taking into account the resulting coupling in the competition between the inter-layer adhesion and elastic strain in the layers, which has been used \cite{Enaldiev_PRL} and tested \cite{Weston2020} earlier in the studies of mesoscale lattice relaxation in TMD bilayers, we derive an effective theory, formulated in terms of the DWN deformations. Finally, we find an analytical solution for such a 'string-like' theory, which has a universal form scaling with the $D/D_*$ ratio. 

%%%%%%%%%%%%%%%%%%%%%%%%%%%%%%%%%%%%%%%%%%%%%%%%%%%%%%%%%%%%%%
\begin{figure}%[!h]
\includegraphics[width=1.0\linewidth]{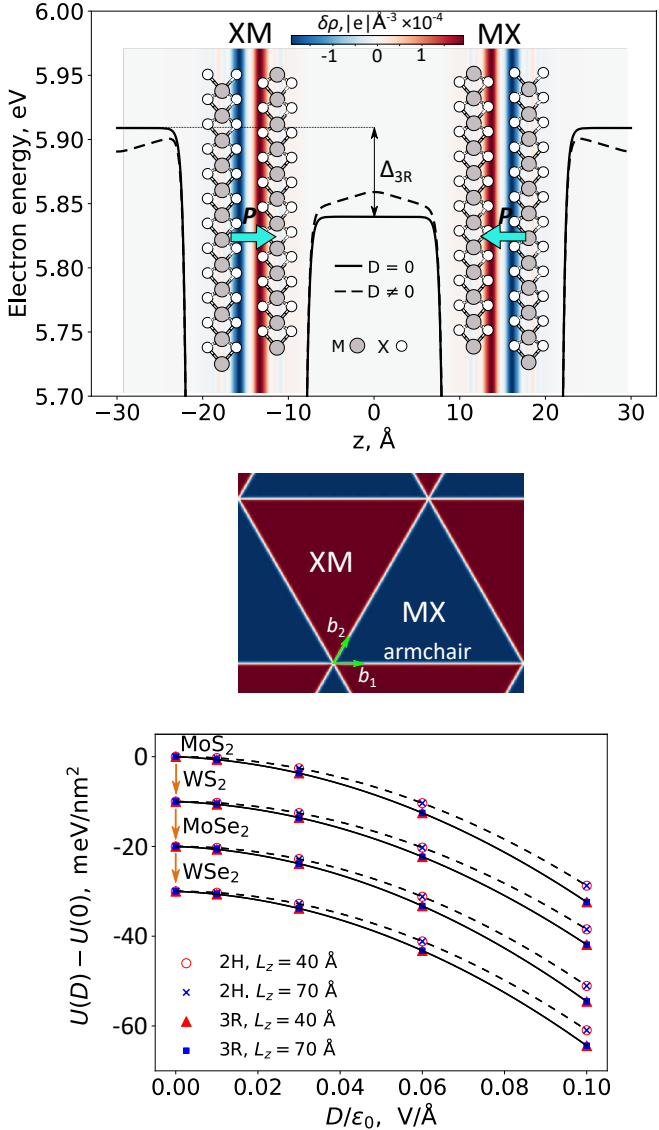}
\caption{\label{Fig1:sketch} Top: Solid (dashed) line shows potential energy of electrons computed in DFT for a double supercell structure of TMD bilayers with 3R stacking (here, MoS$_2$) with and without external displacement field.  A map, superimposed on the lattice structure, visualises FE charge density, $\delta\rho$ (averaged over the unit cell are). The two twin bilayers in the expanded supercell, used in the DFT modelling, represent the actual lattice structure of MX and XM domains in the reconstructed moire superlattice of a TMD bilayer, which at $D=0$ have a triangular shape (middle panel). Bottom: DFT-computed 3R- and 2H-bilayers energy per unit cell as a function of a displacement field. Solid lines are plotted using Eq.(\ref{Eq:DFT_formula}) with parameters listed in Table \ref{tab_polarizability}. } 
\end{figure} 
%%%%%%%%%%%%%%%%%%%%%%%%%%%%%%%%%%%%%%%%%%%%%%%%%%%%%%%%%

{\bf Ab initio modelling of FE bilayers in the out-of-plane electric field.} 
Here, we use two methods to carry out DFT calculations for 3R-bilayers, with the coinciding results.  
In the first method, we use Quantum ESPRESSO (QE) \cite{QE1,QE2} to construct a supercell with a pair of mirror-reflected (MX and XM) bilayers, separated by a large vacuum gap, as shown in Fig. \ref{Fig1:sketch}. 
This choice of the structure eliminates an issue with periodic boundary condition for potential of the FE charges \cite{Ferreira2021}. 
In the second method we construct a single 3R-bilayer and use Coulomb truncation in the out-of-plane direction \cite{CoulombCut} with  QE and dipole-correction with VASP\cite{VASP}. 
An example of the computed charge transfer, $\delta \rho$, between the layers of an individual bilayer (which integral, $\int z \delta\rho (z) dz \equiv P$, determines the areal density of the FE dipole moment) and a potential drop, $\Delta$, across the double charge layer, is shown in Fig. \ref{Fig1:sketch}. To mention, the computed values of $P$ and $\Delta$ are related as $P = \epsilon_0\Delta$, in agreement with the earlier studies of 3D bulk FE materials \cite{Vanderbilt_PRB}. 

The displacement field, accounted for by adding a triangular potential, $-D|z|/\epsilon_0$, to the input pseudo-potentials, produces a shift, $\delta U \equiv U(D)-U(0)$ of the bilayer energy. The latter includes a linear term, accounting for the FE polarization, and a quadratic downward shift due to the dielectric out-of-plane polarizability of the material:
\begin{equation}\label{Eq:DFT_formula}
    \delta U = -\frac{PD}{\chi\epsilon_0}-\frac{\alpha_{zz}^{\rm 3R} D^2}{2\epsilon_0\mathcal{A}}.
\end{equation} 
Here, we parametrise the $P-D$ coupling by a dimensionless parameter $\chi$ and the out-of-plane dielectric polarizability by $\alpha_{zz}^{\rm 3R}$ (where $\mathcal{A}$ is the unit cell area of a TMD monolayer). The DFT-computed energies of bilayers of four different TMDs are shown in Fig. \ref{Fig1:sketch}: in this computation, we used two different supercell periods ($L_z=$40\,\AA\, and $L_z=$70\,\AA, which set the length of the vacuum spacer). Using Eq. (\ref{Eq:DFT_formula}), we find the polarizability values, which are very close to polarizability,  $\alpha^{2H}_{zz}$, computed for the inversion-symmetric 2H bilayer\footnote{For monolayers and 2H-bilayers we used  the Coulomb truncation in the out-of-plane direction\cite{CoulombCut}.}. When recalculated per monolayer, these values also agree with the separately computed monolayer polarizability, $\alpha_{zz}$ ($\alpha_{zz} = \frac12 \alpha^{3R}_{zz} = \frac12 \alpha^{2H}_{zz}$, see in Table \ref{tab_polarizability}). This confirms that the dielectric response of a wide band gap van der Waals materials is determined by the intra-layer polarization of the constituent atoms
\footnote{ Unlike Refs. \cite{santos2013,Laturia2020}, which use $z$-averaged electric field for computation of dielectric permitivities of bilayers, we express the quadratic amendment to the total energy via out-of-plane displacement field, conserving across every cross-section of the structure. This allows us to avoid uncertainties in $\alpha_{zz}^{\rm 3R}$, which may appear at averaging of the electric field in crystals with a few out-of-plane unit cells \cite{Vanderbilt_PRB}. We find that the polarizability, computed for a TMD monolayer, its 2H and 3R bilayer, and thicker 2H films, linearly scales with with the number of layers. This observation contradicts some earlier DFT studies of dielectric susceptibility of TMDs \cite{santos2013,Laturia2020} which claimed a pronounce layer-number-dependence, but agrees with the more recent results \cite{Tian2020} published by some of the authors of Refs \cite{santos2013}. }. 
These values enabled us to estimate the $z$-axis dielectric susceptibility of bulk TMD crystals \cite{Slizovskiy2021_arxiv,Tian2020,Slizovskiy2021}, $\epsilon_{zz}=(1-\frac{\alpha_{zz}}{\mathcal{A}d})^{-1}$, arriving at the values in the range of $\epsilon_{zz} \sim 6-7.5$, listed in Table \ref{tab_polarizability}. 

Despite a substantial polarizability of monolayers, the data in Fig.  \ref{Fig1:sketch} are described well by Eq. (\ref{Eq:DFT_formula}) with $\chi\approx 1$, pointing towards a decoupling of the inter-layer FE charge transfer from the intra-layer dielectric polarizability. Moreover, by comparing the DFT-computed values of the double-layer potential drop $\Delta$ to the FE coupling with the displacement field, $D$, in Eq. \eqref{Eq:interaction_intro}, we find that the linear in $D$ energy shift in MX and XM domains (which have opposite out-of-plane FE polarization, $\pm P$, and voltage drop across the double charge layer, $\pm\Delta$) can be described very well as \footnote{The expression for interaction energy of the FE polarization with displacement field naturally comes when assuming a local dielectric permittivity in a continuum medium approximation. Indeed, suppose the FE charges, with plane-averaged density $\delta\rho(z)$, are placed in the medium with local dielectric permittivity $\epsilon_{zz}(z)$. From the Poisson equation and electro-neutrality condition $\int_{-\infty}^{+\infty}\delta\rho(z)dz=0$ \mbox{$\partial_{z}(\epsilon_{zz}(z)\partial_z\varphi(z))=-\delta\rho(z)/\epsilon_0$}, we express the potential drop across the layer of charges as $\Delta=\int_{-\infty}^{+\infty}\partial_z\varphi(z)dz=-\int_{-\infty}^{+\infty}dz\int_{-\infty}^{z}dz'\delta\rho(z')/\epsilon_{zz}(z)\epsilon_0=\int_{-\infty}^{+\infty}dz\int^{+\infty}_{z}dz'\delta\rho(z')/\epsilon_{zz}(z)\epsilon_0=\int\int_{z<z'}dzdz'\delta\rho(z')/\epsilon_{zz}(z)\epsilon_0$. At the same time, interaction energy of these  charges with uniform external out-of-plane displacement field (related to local electric field as \mbox{$D=\epsilon_0\epsilon_{zz}(z)E(z)$}) reads as $\delta U=-\int_{-\infty}^{+\infty}dz\delta\rho(z)\int_{-\infty}^zdz'D/\epsilon_{zz}(z')\epsilon_0= -D\int_{-\infty}^{+\infty}dz'\int_{-\infty}^{z'}dz\delta\rho(z')/\epsilon_{zz}(z)\epsilon_0=-D\int\int_{z<z'}dzdz'\delta\rho(z')/\epsilon_{zz}(z)\epsilon_0\equiv -D\Delta$. After changing variables, \mbox{$z\leftrightarrow z'$}, at the last step of that calculation, we arrive at the relation in Eq. \eqref{Eq:interaction_intro}.}  
\begin{equation}\label{Eq:interaction_intro}
    \delta U_{\rm MX} = -\delta U_{\rm XM} \approx  - D\Delta.
\end{equation}
Below, we will use this coupling to model tunability of the DWN by displacement field.

%%%%%%%%%%%%%%%%%%%%%%%%%%%%%%%%%%%%

\begin{table}[t]
    \begin{threeparttable}
    	\caption{Dielectric and FE parameters for various TMDs. Left:  polarizability ($\alpha_{zz}$) (used to estimate the dielectric permittivity $\epsilon_{zz}$ of 3D-bulk TMDs), and FE coupling parameter ($\chi$) for various 3R-TMD bilayers\tnote{$*)$}. Right: FE potential drop parameters in Eq. (\ref{Eq:energy_functional_2D}). \label{tab_polarizability}}
    	\begin{tabular}{l|ccccc|ccc}
    		\hline
    		\hline
    	   &  $\alpha^{\rm 3R}_{zz}$  & $\alpha^{\rm 2H}_{zz}$ & $\alpha_{zz}$ & $\epsilon_{zz}$ & $\chi$ & 	$\Delta$, & $\Delta_a$, & $q$,  \\ 
    		 &  \AA$^3$ & \AA$^3$ & \AA$^3$ & &  & mV & mV & nm$^{-1}$  \\
    		\hline 
    		\multirow{2}{*}{MoS$_2$}&   89.89 & 89.80 &44.46 &   6.45 & 1.03 & \multirow{2}{*}{68} & \multirow{2}{*}{16.4} & \multirow{2}{*}{22.152}   \\ 
    		 & 89.93 & 89.85 & 44.48 & 6.48 & 1.02 & & &      \\
    		\multirow{2}{*}{WS$_2$}&  88.43 & 88.37& 43.95&  5.95 & 1.00 & \multirow{2}{*}{62} & \multirow{2}{*}{15} & \multirow{2}{*}{22.598}       \\ 
    		 &  88.43 & 88.37& 43.94 & 5.95 & 1.00 & & &      \\	 
    		\multirow{2}{*}{MoSe$_2$}&  105.32 & 105.22 & 52.29 &  7.65 & 1.05 &\multirow{2}{*}{66} & \multirow{2}{*}{15.7} & \multirow{2}{*}{20.520}    \\ 
    		 & 105.31 & 105.25 & 52.24& 7.67 & 1.07 & & &     \\	
    		\multirow{2}{*}{WSe$_2$}&  104.37 & 104.33 & 52.03 &  7.39 & 1.04 & \multirow{2}{*}{65} & \multirow{2}{*}{15.3} & \multirow{2}{*}{20.953}    \\ 
    		 & 104.32 & 104.28 & 51.97 & 7.32 & 1.07 & & &  \\
    		\hline
    		\hline
    	\end{tabular}

    	\begin{tablenotes}[flushleft]
            \item[$*)$]{\small  For $\alpha^{\rm 3R}_{zz}$, $\chi$ and $\epsilon_{zz}$ we show the values obtained with QE (top row) and - for comparison - with VASP (bottom row). For QE we used full-relativistic ultra-soft pseudo-potentials and a plane-wave cutoff energy of 70 Ry. For VASP, we used PAW full-relativistic pseudo-potentials and 60 Ry cut-off (with a $13 \times 13 \times 1$ $k$-point grid and a Perdew-Burke-Ernzerhof (PBE)\cite{PBE} exchange-correlation functional for both QE and VASP).}
        \end{tablenotes}
    
\end{threeparttable}
\end{table}
%%%%%%%%%%%%%%%%%%%%%%%%%%%%%%%%%%%%%%%%%%%%%%%%%%%%%

\begin{figure}%[t]
\includegraphics[width=1\linewidth]{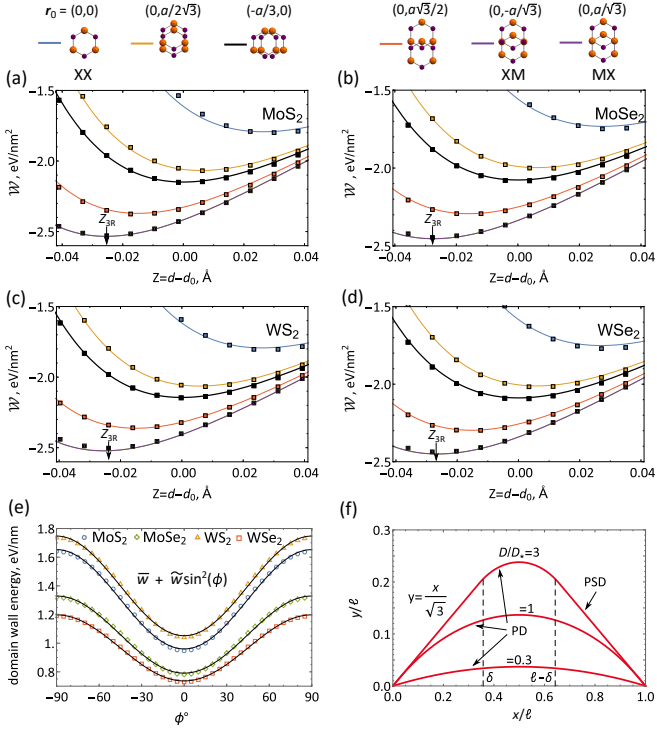}
\caption{\label{Fig:enrg} (a-d) TMD bilayer adhesion energies for various offsets  and interlayer distances, $Z=d-d_0$, counted from the optimal distance, $d_0$, of stacking configurations averaged adhesion energy. $Z_{\rm 3R}$ labels interlayer distance used to evaluate the FE parameters in 3R (MX and XM) TMD bilayers. Details of DFT data (square dots) are described in Ref. \cite{Enaldiev_PRL}.  Insets show stackings with corresponding in-plane offsets between the layers. (e) Orientation dependence of partial dislocation energy per unit length calculated in Ref. \cite{Enaldiev_PRL} and fitted with $\bar{w}+\widetilde{w}\sin^2\phi$, where values of parameters are listed in Table \ref{tab_Fit}. $\phi=0$ corresponds to the armchair direction. (f) Evolution of domain wall shape with displacement field, given by solution of Eq. (\ref{Eq:differential_eq}) for $D<D_*$ and $D\geq D_*$.} 
\end{figure}

{\bf Mesoscale model for lattice reconstruction} is formulated in terms of the bilayer energy dependence on local stacking of the two layers and their strain  \cite{Weston2020,Enaldiev_PRL,Enaldiev_2021}, incorporated via an interlayer offset,  $\bm{r}_0\left(\bm{r}\right)=\theta\hat{z}\times\bm{r}+\bm{u}^{(t)}-\bm{u}^{(b)}$, which varies across the superlattice, as prescribed by a small-angle, $\theta$, misalignment between the top (t) and bottom (b) crystals and their elastic deformations, $\bm{u}^{(t/b)}$. Locally, stacking determines the interlayer distance, which corresponds to the minimum of the stacking-dependent adhesion energy, $\mathcal{W}(\bm{r}_0, d)$, quantified for the four TMDs using DFT modelling and displayed in Fig. \ref{Fig:enrg}. For convenience, the inter-layer distance ($d$) dependence of the computed $\mathcal{W}(\bm{r}_0, d)$ is shown as a function of $Z=d-d_0$, counted from the minimum (at $d=d_0$) of the offset-averaged adhesion energy, which coincides with $\mathcal{W}(\bm{r}_0=[-a/3,0],d)$ (one of stacking configurations shown in Fig. \ref{Fig:enrg}). As a result, energy density, characterizing the relaxation functional, reads:
\begin{align}\label{Eq:energy_functional_2D}
\mathcal{E}= & \sum_{a=t,b}\left[\frac12\lambda\left(u_{ii}^{(a)}\right)^2 + \mu u_{ij}^{(a)}u_{ji}^{(a)}\right] \nonumber \\ 
& + \mathcal{W}\left[\bm{r}_0(\bm{r}),Z\right]  - D\Delta\left[\bm{r}_0(\bm{r}),Z\right];\\% \right\},\\
\mathcal{W}(\bm{r}_0,Z) & =\kappa Z^2+\left(\gamma-\gamma' Z\right)\sum_{n=1,2,3}{\cos\left(\bm{G}_n\cdot\bm{r}_0\right)}. \nonumber  
\end{align}
Here, the first term accounts for strain, $u_{ij}^{(t/b)}=\frac12 \left(\partial_iu_j^{(t/b)}+\partial_ju_i^{(t/b)}\right)$ ($\lambda$ and $\mu$ are the monolayer elastic moduli). The second term describes adhesion energy between top and bottom layers, where $\kappa$ determines curvature of the adhesion energy of 3R-stacked bilayers, which characterises frequency of layer breathing mode \cite{Enaldiev_PRL}, and $\bm{G}_{1,2,3}$ is the first star of reciprocal lattice vectors of TMD monolayer, $\left|\bm{G}_{1,2,3}\right|=4\pi/a\sqrt{3}$ (for details, see Refs. \cite{Weston2020,Enaldiev_PRL,Enaldiev_2021}). The last term is responsible for the energy shift due to the external field. 

The next steps in the analysis require choosing an optimal interlayer distance between the two monolayer, which depends on the encapsulation environment of the bilayer. For a bilayer in vacuum, or encapsulated into a soft flexible matrix with a weaker adhesion to TMD than the the interlayer adhesion $\cal{W}$, we find from Eq. \eqref{Eq:energy_functional_2D} that, locally,  $Z(\bm{r}_0)=\frac{\gamma'}{2\kappa}\sum_{n}\cos(\bm{G}_n\cdot\bm{r}_0)$. For a strong substrate-TMD coupling, the monolayers would remain flat, and the optimal interlayer distance would be $Z=0$ across the entire moi\'re pattern. For either of these two cases, we substitute the corresponding choice of $Z$ in Eq. \eqref{Eq:energy_functional_2D}, and also into the local voltage drop across the double layer of charge (in the last term in Eq. \eqref{Eq:energy_functional_2D}) corresponding to the local stacking configuration, which form has been established earlier \cite{Ferreira2021}:    
$$
\Delta\left[\bm{r}_0,Z(\bm{r}_0)\right]=\Delta_ae^{-qZ\left(\bm{r}_0\right)}\sum_{n=1,2,3}\sin\left(\bm{G}_n\cdot\bm{r}_0\right).
$$
For XM and MX stackings, this gives $\Delta \equiv \Delta(\bm{r}_0^{\rm MX})=-\Delta(\bm{r}_0^{\rm XM})$, which values, taken from the recent {\it ab initio} simulations \cite{Ferreira_APL,Enaldiev_PRL}, are listed in Table \ref{tab_Fit}.

%%%%%%%%%%%%%%%%%%%%%%%%%%%%%%%%%%%%%%%%%%%%%%%%%%%%%%%%%%%%%%%%%%%%%%%%%%%%%%%%%%%
\begin{table*}%[t]
	\caption{Elastic moduli, adhesion and partial dislocation energy density parameters for the studied TMD bilayers. \label{tab_Fit}}
	\begin{tabular}{l|ccccccc|ccc}
		\hline
		\hline
	  &	$\gamma$, & $\gamma'$, & $\kappa$,  &  $\mu$ ,& $\lambda$, & $a$, & $d_0$,  & $\bar{w}$ & $\widetilde{w}$ & $u$ \\ 
		&\mbox{eV/nm$^{2}$} &\mbox{eV/nm$^{3}$} &  \mbox{eV/nm$^{4}$} &  N/m & N/m & nm & nm  & eV/nm & eV/nm & eV/nm \\
		\hline 
		MoS$_2$ & 0.189 &  5.634 & 214 &  70.9 & 83.2 & 0.316 & 0.636 & 0.96 & 0.69 & 2.24 \\ 
		WS$_2$ &  0.212 &  6.327 & 213 &  72.5 & 52.5 &  0.315 & 0.638 & 1.05 & 0.69 & 2.45 \\
		MoSe$_2$ & 0.1975  & 5.707 & 189 & 49.6 &  42.3 & 0.329 & 0.670 & 0.79 & 0.54 & 1.84 \\
		WSe$_2$ & 0.1514 &  4.381 & 190  & 48.4 & 29.7 &  0.328 & 0.671 & 0.74 & 0.46 & 1.69 \\
		\hline
		\hline
	\end{tabular}
\end{table*}

%%%%%%%%%%%%%%%%%%%%%%%%%%%%%%%%%%%%%%%%%%%%%%%%%%%%%%%%%%%%%%%%%%%%%%%%%%%%%%%%%%%%%%%%%
%%%%%%%%%%%%%%%%%%%%%%%%%%%%%%%%%%%%%%%%%%%%%%%%%%%%%%%%%%%%%%%%%%%%%%%%%%%%%%%%%%%%%%%%%

{\bf Network model.} While the minimization of the energy functional in  \eqref{Eq:energy_functional_2D} enables one to find all mesoscale details of the lattice adjustments of the two crystalline plane of the bilayer, it is more practical to use another method for the analysis of small-angle ($\theta\lesssim 0.5^{\circ}$) twisted bilayers. This is because, in the latter case, most of the areas of the moir\'e pattern is occupied by the homogeneous MX and XM stacking domains with a characteristic size of $\ell\gtrsim 50$\,nm, whereas the deformations are concentrated inside narrow domain walls, which are only only few nanometres in width \cite{Enaldiev_PRL,Weston2020}. Then, we define the DWN energy, 
$$\mathcal{E}_{\ell} \equiv \int_{\rm supercell} d^2\bm{r}\left\{\mathcal{E}-\mathcal{W}\left[\bm{r}_0^{\rm MX},Z\left(\bm{r}_0^{\rm MX}\right)\right]\right\},$$ 
by subtracting the energy of a uniform MX-stacked bilayer at $D=0$ from $\cal{E}$ in Eq.  \eqref{Eq:energy_functional_2D}. This reduces the problem to the analysis of $\mathcal{E}_{\ell}$, where we can treat each domain wall as a string, characterised by energy per unit length dependent on a crystallographic orientation of the domain wall axis \cite{Enaldiev_PRL}, with - as shown in Fig. \ref{Fig:enrg}(e) - a pronounced minimum at the armchair direction in the TMD crystal. 

A finite displacement field acts as an external drive for increasing areas of domains with the preferential ferroelectric polarization. This leads to the bending of the partial dislocation (PD) domain walls, hence, increasing their energy due to their elongation and the PD axis deviation from the armchair axis. The exact form on the deformed DWN can be found by solving a 'string theory' model, expressed in terms of a deflection, $y(x)$, of PD segments from the closest armchair direction, formalised using an energy functional,
\begin{widetext}
\begin{equation}\label{Eq:network_Enrg}
     \mathcal{E}_{\ell}[y\left(x\right),\delta]=
     3\int_{\delta}^{\ell-\delta}\left[\left(\bar{w}+\widetilde{w}\frac{y'^2}{1+y'^2}\right)\sqrt{1+y'^2}-2yD\Delta \right]dx + 2\sqrt{3}\left[u-D\Delta\delta\right]\delta. 
 \end{equation}
\end{widetext}
Here, the first term in the integral describes the orientation-dependence of the PD energy, $\bar{w}+\widetilde{w}\sin^2{\phi}$, and its stretching, accounted by a factor $\sqrt{1+{y'}^2}$ with $\phi$ standing for an angle between the dislocation axis and the closest armchair direction in the crystal, so that $\sin^2\phi=y'^2/(1+y'^2)$. Two 'stiffness' parameters, $\bar{w}$ and $\widetilde{w}$, were determined using the data from Ref. \cite{Enaldiev_PRL}, see Table \ref{tab_Fit} for their values for various TMDs. The second term in the integral stands for the energy gain from a bigger area of the energetically preferential FE polarization domain. The last two terms in $\mathcal{E}_{\ell}$ accounts for two PDs merging, near each DWN node, into streaks of PSDs with a length $\mathcal{L}=\delta/\sqrt{3}$ (projected onto $0\leq x\leq \delta$ and $\delta\leq x \leq\delta-\ell$ intervals), which energetically preferable orientation is along zigzag axis in the crystal and energy per unit length is \mbox{$u$} (Table \ref{tab_Fit}).

Using variational principle, we obtain an equation for the the shape of each string, 
\begin{equation}\label{Eq:differential_eq}
\begin{split}
y^{\prime\prime}&=-\frac{2D\Delta \left(1+y'^2\right)^\frac{5}{2}}{\bar{w}+2\widetilde{w}+(\bar{w}-\widetilde{w})y'^2}, \\
y&(\delta)=y(\ell-\delta)=\frac{\delta}{\sqrt{3}}, 
\end{split}
\end{equation}
where $\delta$ appears as an additional variable. Note that zero values of $y(x)$ at the network nodes requires vanishing of derivative in the middle of interval, i.e. providing a symmetrical shape of the PD, with $y'(\ell/2)=0$.

For a small $D$, energetically favourable domains grow due to bending of PDs, with their ends  fixed at network nodes, \mbox{$\delta=0$}. This elastic stretching has no threshold in $D$ and the string form is described by \footnote{At small $D$ (when \mbox{$y'^2\ll 1$}) approximate solution of Eq. \eqref{Eq:differential_eq} is given by parabola: \mbox{$y(x)=2D\Delta (\ell -x)x/(\bar{w}+2\widetilde{w})$.}},
\begin{equation}\label{Eq:exact_small_D}
 y(x)=\int_{0}^{x}\frac{f(x')}{\sqrt{1-f^2(x')}}dx', \quad 0\leq x\leq\ell,    
\end{equation}
where $f(x)$ is a real root of
\begin{equation}\label{Eq:polynomial}
    f^3-\left(\frac{\bar{w}}{\widetilde{w}}+2\right)f - 2\frac{D\Delta}{\widetilde{w}}\left(x-\frac{\ell}{2}\right) = 0.
\end{equation}
Note that \mbox{$f(\ell/2)=0$} and that $f(x)\equiv 0$ for $D=0$.

\begin{figure}
\includegraphics[width=1\linewidth]{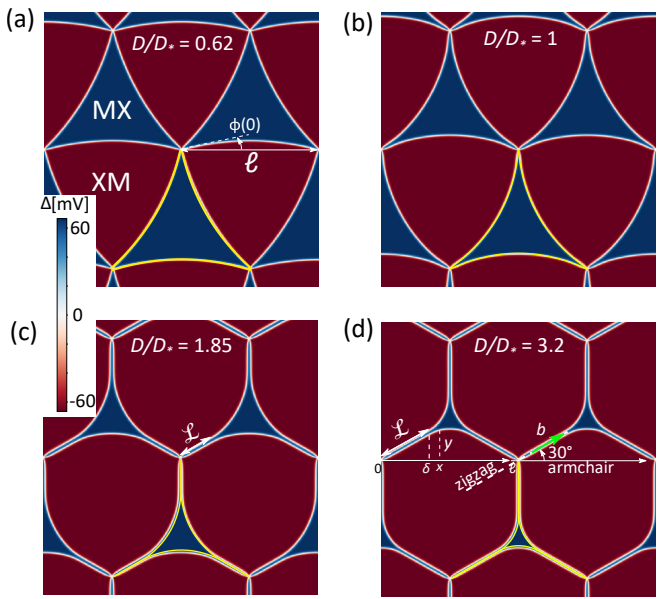}
\caption{\label{Fig:network} (a-d) Maps of the interlayer potential drop, $\Delta$ in MoS$_2$ bilayer with a $\theta = 0.1^\circ$  twist, computed by the minimisation of energy in Eq. (\ref{Eq:energy_functional_2D}) for various values of $D$ below and above the threshold. These maps display the deformation of DWN from a triangular web of partial dislocations - see Figs. \ref{Fig0} and \ref{Fig1:sketch}, with a $\ell \approx 180$ nm period. Yellow lines are the streaks of PSD and arcs of PD described by Eqs. (\ref{Eq:exact_small_D})-(\ref{Eq:universal_sol}).} 
\end{figure}

Bending, described by Eq. \eqref{Eq:exact_small_D}, takes a new qualitative form  
after each pair of PDs near the network nodes ($x=0;\ x=\ell$) touch each other, see Fig. \ref{Fig:network}(j). This condition corresponds to  $\phi^{\circ}(0)=-\phi^{\circ}(l)=30^{\circ}$ (i.e., $y'\left(0\right)=-y'\left(\ell\right)=1/\sqrt{3}$), which, together with Eq. (\ref{Eq:exact_small_D}) determines a threshold displacement field, 
\begin{align}\label{Eq:D_threshold}
    D_\ast&=\epsilon_0\frac{\mathcal{V}}{\ell}, \\ 
    \mathcal{V}= &\frac{\left(\frac{\bar{w}}{2}+\frac{7\widetilde{w}}{8}\right)}{\epsilon_0\Delta}
    \Rightarrow
    \begin{cases}
     \mathcal{V}_{\rm MoS_2}=287\, {\rm V},\mathcal{V}_{\rm WS_2}=310\, {\rm V}, \\
     \mathcal{V}_{\rm MoSe_2}=236\, {\rm V},\mathcal{V}_{\rm WSe_2}=214\, {\rm V}.
    \end{cases}\nonumber
\end{align}

It is important to note that pairs of touching PDs cannot annihilate, as a sum of their Burgers vectors \mbox{$\bm{b}_1+\bm{b}_2=\bm{b}$} (\mbox{$\bm{b}_1=a(1/\sqrt{3},0)$}, \mbox{$\bm{b}_2=a(1/2\sqrt{3},1/2)$}, \mbox{$\bm{b}=a(\sqrt{3}/2,1/2)$}) corresponds to the Burgers vector ($|\bm{b}|=a$) of a PSD, required to maintain an overall twist between two monolayers in the bilayer. This means that they actually form streaks of PSDs, which lengths grow with a further increase of the electric drive, as   
\begin{equation}\label{Eq:PSD_length}
    \mathcal{L}=\frac{\ell}{\sqrt{3}}\left(1-\frac{D_*}{D}\right),
\end{equation}
The latter scaling law follows from the solution of Eq. \eqref{Eq:differential_eq} on the interval \mbox{$\delta\leq x \leq\ell-\delta$} with the boundary conditions, \mbox{$y^\prime\left(\delta\right)=-y^\prime\left(\ell-\delta\right)=1/\sqrt3$}, where $\delta=\sqrt{3}\mathcal{L}/2$ is a projection of $\mathcal{L}$ onto the $x$-axis. Hence, above the threshold, DWN is composed of (A) PSD streaks near each network node aligned with zigzag directions in TMD, and separating the adjacent expanded energetically favourable 3R-domains and (B) PD arcs splitting up at the PSD ends, \mbox{$x=\delta$} and \mbox{$x=\ell-\delta$}, and touching each other at the splitting points.

To find the shape of the remaining PD arcs for any $D>D_*$, we rewrite Eq. \eqref{Eq:differential_eq} in dimensionless coordinates, $\widetilde{x}$ and $\widetilde{y}$ defined by \mbox{$x=\ell\widetilde{x}D_\ast/D+\delta$}, \mbox{$y=\ell\widetilde{y}D_\ast/D+\delta/\sqrt3$},
\begin{align}\label{Eq:scaled_differential}
    \widetilde{y}''=&-\frac{\bar{w}+\frac{7}{4}\widetilde{w}}{\bar{w}+2\widetilde{w}}\frac{\left(1+\widetilde{y}'^2\right)^\frac{5}{2}}{1+\frac{\bar{w}-\widetilde{w}}{\bar{w}+2\widetilde{w}}\widetilde{y}'^2},\quad 0\leq\widetilde{x}\leq 1;   \nonumber \\
    & \widetilde{y}'(0) = -\widetilde{y}'(1)=\frac{1}{\sqrt{3}}. \nonumber
\end{align}
Its solution has the same structure for any $D\geq D_*$,
\begin{eqnarray}\label{Eq:universal_sol}
\widetilde{y}\left(\widetilde{x}\right)=\int_{0}^{\widetilde{x}}{\frac{\widetilde{f}\left(\widetilde{x}'\right)}{\sqrt{1-\widetilde{f}^2\left(\widetilde{x}'\right)}}d\widetilde{x}'},  \qquad\widetilde{f}(0)=\frac{1}{2}; \\
 \widetilde{f}^3-\left(\frac{\bar{w}}{\widetilde{w}}+2\right)\widetilde{f} - \left(\frac{\bar{w}}{\widetilde{w}}+\frac{7}{4}\right)\left(\widetilde{x}-\frac{1}{2}\right) = 0, \nonumber
\end{eqnarray}
describing the universal shape of partial dislocation arcs, scaled by the ratio $D_*/D$, as shown in Fig. \ref{Fig:enrg}(f).

Finally, to validate the accuracy of the obtained analytical solution, obtained using the network model, we also performed the mesoscale lattice relaxation for a MoS$_2$ bilayer with a $0.1^\circ$ interlayer twist (corresponding to $\ell \approx 180$\,nm), by minimising the full energy functional in Eq. (\ref{Eq:energy_functional_2D}). For this, we analysed the Euler-Lagrange equations for the displacement fields $\bm{u}^{(t/b)}$, obtained by applying variational principle to $\mathcal{E} \left[\bm{u}^{(t/b)} \right]$. We solved those numerically, seeking for periodic (with the the period $\ell$) solutions on a sufficiently dense grid using interior point method implemented in GEKKO Optimization Suite package \cite{gekko}. The computed fields were used to obtain the local values of the inter-layer offset, $\bm{r}_0\left(\bm{r}\right)=\theta\hat{z}\times\bm{r}+\bm{u}^{(t)}-\bm{u}^{(b)}$, which we substituted into Eqs. (\ref{Eq:energy_functional_2D}) to find the interlayer distance, $Z$, and to map the interlayer potential drop across the domain structure. The results for the latter quantity are shown in Fig. \ref{Fig:network} for various values of displacement field, displaying a close agreement with the analytical solution of the network model shown by yellow lines. 

%%%%%%%%%%%%%%%%%%%%%%%%%%%%%%%%%%%%%%%%%%%%%%%%%%%%%%%%%%%%
%%%%%%%%%%%%%%%%%%%%%%%%%%%%%%%%%%%%%%%%%%%%%%%%%%%%%%%%%%%%%%

To conclude, the developed effective network model gives an efficient description of the long-period domain structure in small-angle-twisted TMD bilayers with ferroelectric interface and its deformation by an out-of-plane electric field. Its analytical solution in Eqs. (\ref{Eq:exact_small_D}) - (\ref{Eq:universal_sol}) gives a simple tool for the interpretation of experimentally measured variations of the domains' shapes, such as in the recently studied bilayers with ferroelectric properties \cite{yasuda2021,woods2021,stern2021,weston2021,wang2021}.

{\bf Acknowledgments.}
We thank Andre Geim, Roman Gorbachev, Philip Kim, and David Vanderbilt for useful discussions. This study has been supported by the European Graphene Flagship Core3 Project, EU Quantum Flagship Project 2D-SIPC, and EPSRC grants P/W006502/1, EP/V007033/1, EP/S030719/1. The reported {\it ab initio} simulations were performed using EPSRC ARCHER facility and the University of Manchester CSF.

\bibliography{refer}

\begin{thebibliography}{56}%
\makeatletter
\providecommand \@ifxundefined [1]{%
 \@ifx{#1\undefined}
}%
\providecommand \@ifnum [1]{%
 \ifnum #1\expandafter \@firstoftwo
 \else \expandafter \@secondoftwo
 \fi
}%
\providecommand \@ifx [1]{%
 \ifx #1\expandafter \@firstoftwo
 \else \expandafter \@secondoftwo
 \fi
}%
\providecommand \natexlab [1]{#1}%
\providecommand \enquote  [1]{``#1''}%
\providecommand \bibnamefont  [1]{#1}%
\providecommand \bibfnamefont [1]{#1}%
\providecommand \citenamefont [1]{#1}%
\providecommand \href@noop [0]{\@secondoftwo}%
\providecommand \href [0]{\begingroup \@sanitize@url \@href}%
\providecommand \@href[1]{\@@startlink{#1}\@@href}%
\providecommand \@@href[1]{\endgroup#1\@@endlink}%
\providecommand \@sanitize@url [0]{\catcode `\\12\catcode `\$12\catcode
  `\&12\catcode `\#12\catcode `\^12\catcode `\_12\catcode `\%12\relax}%
\providecommand \@@startlink[1]{}%
\providecommand \@@endlink[0]{}%
\providecommand \url  [0]{\begingroup\@sanitize@url \@url }%
\providecommand \@url [1]{\endgroup\@href {#1}{\urlprefix }}%
\providecommand \urlprefix  [0]{URL }%
\providecommand \Eprint [0]{\href }%
\providecommand \doibase [0]{https://doi.org/}%
\providecommand \selectlanguage [0]{\@gobble}%
\providecommand \bibinfo  [0]{\@secondoftwo}%
\providecommand \bibfield  [0]{\@secondoftwo}%
\providecommand \translation [1]{[#1]}%
\providecommand \BibitemOpen [0]{}%
\providecommand \bibitemStop [0]{}%
\providecommand \bibitemNoStop [0]{.\EOS\space}%
\providecommand \EOS [0]{\spacefactor3000\relax}%
\providecommand \BibitemShut  [1]{\csname bibitem#1\endcsname}%
\let\auto@bib@innerbib\@empty
%</preamble>
\bibitem [{\citenamefont {Cao}\ \emph {et~al.}(2018{\natexlab{a}})\citenamefont
  {Cao}, \citenamefont {Fatemi}, \citenamefont {Fang}, \citenamefont
  {Watanabe}, \citenamefont {Taniguchi}, \citenamefont {Kaxiras},\ and\
  \citenamefont {Jarillo-Herrero}}]{cao2018unconventional}%
  \BibitemOpen
  \bibfield  {author} {\bibinfo {author} {\bibfnamefont {Y.}~\bibnamefont
  {Cao}}, \bibinfo {author} {\bibfnamefont {V.}~\bibnamefont {Fatemi}},
  \bibinfo {author} {\bibfnamefont {S.}~\bibnamefont {Fang}}, \bibinfo {author}
  {\bibfnamefont {K.}~\bibnamefont {Watanabe}}, \bibinfo {author}
  {\bibfnamefont {T.}~\bibnamefont {Taniguchi}}, \bibinfo {author}
  {\bibfnamefont {E.}~\bibnamefont {Kaxiras}},\ and\ \bibinfo {author}
  {\bibfnamefont {P.}~\bibnamefont {Jarillo-Herrero}},\ }\bibfield  {title}
  {\bibinfo {title} {Unconventional superconductivity in magic-angle graphene
  superlattices},\ }\href {https://doi.org/10.1038/nature26160} {\bibfield
  {journal} {\bibinfo  {journal} {Nature}\ }\textbf {\bibinfo {volume} {556}},\
  \bibinfo {pages} {43} (\bibinfo {year} {2018}{\natexlab{a}})}\BibitemShut
  {NoStop}%
\bibitem [{\citenamefont {Cao}\ \emph {et~al.}(2018{\natexlab{b}})\citenamefont
  {Cao}, \citenamefont {Fatemi}, \citenamefont {Demir}, \citenamefont {Fang},
  \citenamefont {Tomarken}, \citenamefont {Luo}, \citenamefont
  {Sanchez-Yamagishi}, \citenamefont {Watanabe}, \citenamefont {Taniguchi},
  \citenamefont {Kaxiras},\ and\ \citenamefont {et~al}}]{cao2018correlated}%
  \BibitemOpen
  \bibfield  {author} {\bibinfo {author} {\bibfnamefont {Y.}~\bibnamefont
  {Cao}}, \bibinfo {author} {\bibfnamefont {V.}~\bibnamefont {Fatemi}},
  \bibinfo {author} {\bibfnamefont {A.}~\bibnamefont {Demir}}, \bibinfo
  {author} {\bibfnamefont {S.}~\bibnamefont {Fang}}, \bibinfo {author}
  {\bibfnamefont {S.~L.}\ \bibnamefont {Tomarken}}, \bibinfo {author}
  {\bibfnamefont {J.~Y.}\ \bibnamefont {Luo}}, \bibinfo {author} {\bibfnamefont
  {J.~D.}\ \bibnamefont {Sanchez-Yamagishi}}, \bibinfo {author} {\bibfnamefont
  {K.}~\bibnamefont {Watanabe}}, \bibinfo {author} {\bibfnamefont
  {T.}~\bibnamefont {Taniguchi}}, \bibinfo {author} {\bibfnamefont
  {E.}~\bibnamefont {Kaxiras}},\ and\ \bibinfo {author} {\bibnamefont
  {et~al}},\ }\bibfield  {title} {\bibinfo {title} {Correlated insulator
  behaviour at half-filling in magic-angle graphene superlattices},\ }\href
  {https://doi.org/10.1038/nature26154} {\bibfield  {journal} {\bibinfo
  {journal} {Nature}\ }\textbf {\bibinfo {volume} {556}},\ \bibinfo {pages}
  {80} (\bibinfo {year} {2018}{\natexlab{b}})}\BibitemShut {NoStop}%
\bibitem [{\citenamefont {{Yankowitz}}\ \emph {et~al.}(2019)\citenamefont
  {{Yankowitz}}, \citenamefont {{Chen}}, \citenamefont {{Polshyn}},
  \citenamefont {{Zhang}}, \citenamefont {{Watanabe}}, \citenamefont
  {{Taniguchi}}, \citenamefont {{Graf}}, \citenamefont {{Young}},\ and\
  \citenamefont {{Dean}}}]{Yankowitz2019}%
  \BibitemOpen
  \bibfield  {author} {\bibinfo {author} {\bibfnamefont {M.}~\bibnamefont
  {{Yankowitz}}}, \bibinfo {author} {\bibfnamefont {S.}~\bibnamefont {{Chen}}},
  \bibinfo {author} {\bibfnamefont {H.}~\bibnamefont {{Polshyn}}}, \bibinfo
  {author} {\bibfnamefont {Y.}~\bibnamefont {{Zhang}}}, \bibinfo {author}
  {\bibfnamefont {K.}~\bibnamefont {{Watanabe}}}, \bibinfo {author}
  {\bibfnamefont {T.}~\bibnamefont {{Taniguchi}}}, \bibinfo {author}
  {\bibfnamefont {D.}~\bibnamefont {{Graf}}}, \bibinfo {author} {\bibfnamefont
  {A.~F.}\ \bibnamefont {{Young}}},\ and\ \bibinfo {author} {\bibfnamefont
  {C.~R.}\ \bibnamefont {{Dean}}},\ }\bibfield  {title} {\bibinfo {title}
  {{Tuning superconductivity in twisted bilayer graphene}},\ }\href
  {https://doi.org/10.1126/science.aav1910} {\bibfield  {journal} {\bibinfo
  {journal} {Science}\ }\textbf {\bibinfo {volume} {363}},\ \bibinfo {pages}
  {1059} (\bibinfo {year} {2019})}\BibitemShut {NoStop}%
\bibitem [{\citenamefont {Lu}\ \emph {et~al.}(2019)\citenamefont {Lu},
  \citenamefont {Stepanov}, \citenamefont {Yang}, \citenamefont {Xie},
  \citenamefont {Aamir}, \citenamefont {Das}, \citenamefont {Urgell},
  \citenamefont {Watanabe}, \citenamefont {Taniguchi}, \citenamefont {Zhang}
  \emph {et~al.}}]{lu2019superconductors}%
  \BibitemOpen
  \bibfield  {author} {\bibinfo {author} {\bibfnamefont {X.}~\bibnamefont
  {Lu}}, \bibinfo {author} {\bibfnamefont {P.}~\bibnamefont {Stepanov}},
  \bibinfo {author} {\bibfnamefont {W.}~\bibnamefont {Yang}}, \bibinfo {author}
  {\bibfnamefont {M.}~\bibnamefont {Xie}}, \bibinfo {author} {\bibfnamefont
  {M.~A.}\ \bibnamefont {Aamir}}, \bibinfo {author} {\bibfnamefont
  {I.}~\bibnamefont {Das}}, \bibinfo {author} {\bibfnamefont {C.}~\bibnamefont
  {Urgell}}, \bibinfo {author} {\bibfnamefont {K.}~\bibnamefont {Watanabe}},
  \bibinfo {author} {\bibfnamefont {T.}~\bibnamefont {Taniguchi}}, \bibinfo
  {author} {\bibfnamefont {G.}~\bibnamefont {Zhang}}, \emph {et~al.},\
  }\bibfield  {title} {\bibinfo {title} {Superconductors, orbital magnets and
  correlated states in magic-angle bilayer graphene},\ }\href
  {https://doi.org/10.1038/s41586-019-1695-0} {\bibfield  {journal} {\bibinfo
  {journal} {Nature}\ }\textbf {\bibinfo {volume} {574}},\ \bibinfo {pages}
  {653} (\bibinfo {year} {2019})}\BibitemShut {NoStop}%
\bibitem [{\citenamefont {Sharpe}\ \emph {et~al.}(2019)\citenamefont {Sharpe},
  \citenamefont {Fox}, \citenamefont {Barnard}, \citenamefont {Finney},
  \citenamefont {Watanabe}, \citenamefont {Taniguchi}, \citenamefont
  {Kastner},\ and\ \citenamefont {Goldhaber-Gordon}}]{sharpe2019emergent}%
  \BibitemOpen
  \bibfield  {author} {\bibinfo {author} {\bibfnamefont {A.~L.}\ \bibnamefont
  {Sharpe}}, \bibinfo {author} {\bibfnamefont {E.~J.}\ \bibnamefont {Fox}},
  \bibinfo {author} {\bibfnamefont {A.~W.}\ \bibnamefont {Barnard}}, \bibinfo
  {author} {\bibfnamefont {J.}~\bibnamefont {Finney}}, \bibinfo {author}
  {\bibfnamefont {K.}~\bibnamefont {Watanabe}}, \bibinfo {author}
  {\bibfnamefont {T.}~\bibnamefont {Taniguchi}}, \bibinfo {author}
  {\bibfnamefont {M.}~\bibnamefont {Kastner}},\ and\ \bibinfo {author}
  {\bibfnamefont {D.}~\bibnamefont {Goldhaber-Gordon}},\ }\bibfield  {title}
  {\bibinfo {title} {Emergent ferromagnetism near three-quarters filling in
  twisted bilayer graphene},\ }\href {https://doi.org/10.1126/science.aaw3780}
  {\bibfield  {journal} {\bibinfo  {journal} {Science}\ }\textbf {\bibinfo
  {volume} {365}},\ \bibinfo {pages} {605} (\bibinfo {year}
  {2019})}\BibitemShut {NoStop}%
\bibitem [{\citenamefont {Cao}\ \emph {et~al.}(2021)\citenamefont {Cao},
  \citenamefont {Rodan-Legrain}, \citenamefont {Park}, \citenamefont {Yuan},
  \citenamefont {Watanabe}, \citenamefont {Taniguchi}, \citenamefont
  {Fernandes}, \citenamefont {Fu},\ and\ \citenamefont
  {Jarillo-Herrero}}]{cao2021nematicity}%
  \BibitemOpen
  \bibfield  {author} {\bibinfo {author} {\bibfnamefont {Y.}~\bibnamefont
  {Cao}}, \bibinfo {author} {\bibfnamefont {D.}~\bibnamefont {Rodan-Legrain}},
  \bibinfo {author} {\bibfnamefont {J.~M.}\ \bibnamefont {Park}}, \bibinfo
  {author} {\bibfnamefont {N.~F.}\ \bibnamefont {Yuan}}, \bibinfo {author}
  {\bibfnamefont {K.}~\bibnamefont {Watanabe}}, \bibinfo {author}
  {\bibfnamefont {T.}~\bibnamefont {Taniguchi}}, \bibinfo {author}
  {\bibfnamefont {R.~M.}\ \bibnamefont {Fernandes}}, \bibinfo {author}
  {\bibfnamefont {L.}~\bibnamefont {Fu}},\ and\ \bibinfo {author}
  {\bibfnamefont {P.}~\bibnamefont {Jarillo-Herrero}},\ }\bibfield  {title}
  {\bibinfo {title} {Nematicity and competing orders in superconducting
  magic-angle graphene},\ }\href {https://doi.org/10.1126/science.abc2836}
  {\bibfield  {journal} {\bibinfo  {journal} {Science}\ }\textbf {\bibinfo
  {volume} {372}},\ \bibinfo {pages} {264} (\bibinfo {year}
  {2021})}\BibitemShut {NoStop}%
\bibitem [{\citenamefont {Kerelsky}\ \emph {et~al.}(2021)\citenamefont
  {Kerelsky}, \citenamefont {Rubio-Verd{\'u}}, \citenamefont {Xian},
  \citenamefont {Kennes}, \citenamefont {Halbertal}, \citenamefont {Finney},
  \citenamefont {Song}, \citenamefont {Turkel}, \citenamefont {Wang},
  \citenamefont {Watanabe}, \citenamefont {Taniguchi}, \citenamefont {Hone},
  \citenamefont {Dean}, \citenamefont {Basov}, \citenamefont {Rubio},\ and\
  \citenamefont {Pasupathy}}]{Kerelskye2021}%
  \BibitemOpen
  \bibfield  {author} {\bibinfo {author} {\bibfnamefont {A.}~\bibnamefont
  {Kerelsky}}, \bibinfo {author} {\bibfnamefont {C.}~\bibnamefont
  {Rubio-Verd{\'u}}}, \bibinfo {author} {\bibfnamefont {L.}~\bibnamefont
  {Xian}}, \bibinfo {author} {\bibfnamefont {D.~M.}\ \bibnamefont {Kennes}},
  \bibinfo {author} {\bibfnamefont {D.}~\bibnamefont {Halbertal}}, \bibinfo
  {author} {\bibfnamefont {N.}~\bibnamefont {Finney}}, \bibinfo {author}
  {\bibfnamefont {L.}~\bibnamefont {Song}}, \bibinfo {author} {\bibfnamefont
  {S.}~\bibnamefont {Turkel}}, \bibinfo {author} {\bibfnamefont
  {L.}~\bibnamefont {Wang}}, \bibinfo {author} {\bibfnamefont {K.}~\bibnamefont
  {Watanabe}}, \bibinfo {author} {\bibfnamefont {T.}~\bibnamefont {Taniguchi}},
  \bibinfo {author} {\bibfnamefont {J.}~\bibnamefont {Hone}}, \bibinfo {author}
  {\bibfnamefont {C.}~\bibnamefont {Dean}}, \bibinfo {author} {\bibfnamefont
  {D.~N.}\ \bibnamefont {Basov}}, \bibinfo {author} {\bibfnamefont
  {A.}~\bibnamefont {Rubio}},\ and\ \bibinfo {author} {\bibfnamefont {A.~N.}\
  \bibnamefont {Pasupathy}},\ }\bibfield  {title} {\bibinfo {title}
  {Moir{\'e}less correlations in abca graphene},\ }\href
  {https://www.pnas.org/content/118/4/e2017366118} {\bibfield  {journal}
  {\bibinfo  {journal} {PNAS}\ }\textbf {\bibinfo {volume} {118}} (\bibinfo
  {year} {2021})}\BibitemShut {NoStop}%
\bibitem [{\citenamefont {Xu}\ \emph {et~al.}(2019)\citenamefont {Xu},
  \citenamefont {Berdyugin}, \citenamefont {Kumaravadivel}, \citenamefont
  {Guinea}, \citenamefont {Krishna~Kumar}, \citenamefont {Bandurin},
  \citenamefont {Morozov}, \citenamefont {Kuang}, \citenamefont {Tsim},
  \citenamefont {Liu}, \citenamefont {Edgar}, \citenamefont {Grigorieva},
  \citenamefont {Fal'ko}, \citenamefont {Kim},\ and\ \citenamefont
  {Geim}}]{Xu2019}%
  \BibitemOpen
  \bibfield  {author} {\bibinfo {author} {\bibfnamefont {S.~G.}\ \bibnamefont
  {Xu}}, \bibinfo {author} {\bibfnamefont {A.~I.}\ \bibnamefont {Berdyugin}},
  \bibinfo {author} {\bibfnamefont {P.}~\bibnamefont {Kumaravadivel}}, \bibinfo
  {author} {\bibfnamefont {F.}~\bibnamefont {Guinea}}, \bibinfo {author}
  {\bibfnamefont {R.}~\bibnamefont {Krishna~Kumar}}, \bibinfo {author}
  {\bibfnamefont {D.~A.}\ \bibnamefont {Bandurin}}, \bibinfo {author}
  {\bibfnamefont {S.~V.}\ \bibnamefont {Morozov}}, \bibinfo {author}
  {\bibfnamefont {W.}~\bibnamefont {Kuang}}, \bibinfo {author} {\bibfnamefont
  {B.}~\bibnamefont {Tsim}}, \bibinfo {author} {\bibfnamefont {S.}~\bibnamefont
  {Liu}}, \bibinfo {author} {\bibfnamefont {J.~H.}\ \bibnamefont {Edgar}},
  \bibinfo {author} {\bibfnamefont {I.~V.}\ \bibnamefont {Grigorieva}},
  \bibinfo {author} {\bibfnamefont {V.~I.}\ \bibnamefont {Fal'ko}}, \bibinfo
  {author} {\bibfnamefont {M.}~\bibnamefont {Kim}},\ and\ \bibinfo {author}
  {\bibfnamefont {A.~K.}\ \bibnamefont {Geim}},\ }\bibfield  {title} {\bibinfo
  {title} {Giant oscillations in a triangular network of one-dimensional states
  in marginally twisted graphene},\ }\href
  {https://doi.org/10.1038/s41467-019-11971-7} {\bibfield  {journal} {\bibinfo
  {journal} {Nature Communications}\ }\textbf {\bibinfo {volume} {10}},\
  \bibinfo {pages} {4008} (\bibinfo {year} {2019})}\BibitemShut {NoStop}%
\bibitem [{\citenamefont {Gadelha}\ \emph {et~al.}(2021)\citenamefont
  {Gadelha}, \citenamefont {Ohlberg}, \citenamefont {Rabelo}, \citenamefont
  {Neto}, \citenamefont {Vasconcelos}, \citenamefont {Campos}, \citenamefont
  {Lemos}, \citenamefont {Ornelas}, \citenamefont {Miranda}, \citenamefont
  {Nadas}, \citenamefont {Santana}, \citenamefont {Watanabe}, \citenamefont
  {Taniguchi}, \citenamefont {van Troeye}, \citenamefont {Lamparski},
  \citenamefont {Meunier}, \citenamefont {Nguyen}, \citenamefont {Paszko},
  \citenamefont {Charlier}, \citenamefont {Campos}, \citenamefont
  {Can{\c{c}}ado}, \citenamefont {Medeiros-Ribeiro},\ and\ \citenamefont
  {Jorio}}]{Gadelha2021}%
  \BibitemOpen
  \bibfield  {author} {\bibinfo {author} {\bibfnamefont {A.~C.}\ \bibnamefont
  {Gadelha}}, \bibinfo {author} {\bibfnamefont {D.~A.~A.}\ \bibnamefont
  {Ohlberg}}, \bibinfo {author} {\bibfnamefont {C.}~\bibnamefont {Rabelo}},
  \bibinfo {author} {\bibfnamefont {E.~G.~S.}\ \bibnamefont {Neto}}, \bibinfo
  {author} {\bibfnamefont {T.~L.}\ \bibnamefont {Vasconcelos}}, \bibinfo
  {author} {\bibfnamefont {J.~L.}\ \bibnamefont {Campos}}, \bibinfo {author}
  {\bibfnamefont {J.~S.}\ \bibnamefont {Lemos}}, \bibinfo {author}
  {\bibfnamefont {V.}~\bibnamefont {Ornelas}}, \bibinfo {author} {\bibfnamefont
  {D.}~\bibnamefont {Miranda}}, \bibinfo {author} {\bibfnamefont
  {R.}~\bibnamefont {Nadas}}, \bibinfo {author} {\bibfnamefont {F.~C.}\
  \bibnamefont {Santana}}, \bibinfo {author} {\bibfnamefont {K.}~\bibnamefont
  {Watanabe}}, \bibinfo {author} {\bibfnamefont {T.}~\bibnamefont {Taniguchi}},
  \bibinfo {author} {\bibfnamefont {B.}~\bibnamefont {van Troeye}}, \bibinfo
  {author} {\bibfnamefont {M.}~\bibnamefont {Lamparski}}, \bibinfo {author}
  {\bibfnamefont {V.}~\bibnamefont {Meunier}}, \bibinfo {author} {\bibfnamefont
  {V.-H.}\ \bibnamefont {Nguyen}}, \bibinfo {author} {\bibfnamefont
  {D.}~\bibnamefont {Paszko}}, \bibinfo {author} {\bibfnamefont {J.-C.}\
  \bibnamefont {Charlier}}, \bibinfo {author} {\bibfnamefont {L.~C.}\
  \bibnamefont {Campos}}, \bibinfo {author} {\bibfnamefont {L.~G.}\
  \bibnamefont {Can{\c{c}}ado}}, \bibinfo {author} {\bibfnamefont
  {G.}~\bibnamefont {Medeiros-Ribeiro}},\ and\ \bibinfo {author} {\bibfnamefont
  {A.}~\bibnamefont {Jorio}},\ }\bibfield  {title} {\bibinfo {title}
  {Localization of lattice dynamics in low-angle twisted bilayer graphene},\
  }\href {https://doi.org/10.1038/s41586-021-03252-5} {\bibfield  {journal}
  {\bibinfo  {journal} {Nature}\ }\textbf {\bibinfo {volume} {590}},\ \bibinfo
  {pages} {405} (\bibinfo {year} {2021})}\BibitemShut {NoStop}%
\bibitem [{\citenamefont {Kazmierczak}\ \emph {et~al.}(2021)\citenamefont
  {Kazmierczak}, \citenamefont {Van~Winkle}, \citenamefont {Ophus},
  \citenamefont {Bustillo}, \citenamefont {Carr}, \citenamefont {Brown},
  \citenamefont {Ciston}, \citenamefont {Taniguchi}, \citenamefont {Watanabe},\
  and\ \citenamefont {Bediako}}]{Kazmierczak2021}%
  \BibitemOpen
  \bibfield  {author} {\bibinfo {author} {\bibfnamefont {N.~P.}\ \bibnamefont
  {Kazmierczak}}, \bibinfo {author} {\bibfnamefont {M.}~\bibnamefont
  {Van~Winkle}}, \bibinfo {author} {\bibfnamefont {C.}~\bibnamefont {Ophus}},
  \bibinfo {author} {\bibfnamefont {K.~C.}\ \bibnamefont {Bustillo}}, \bibinfo
  {author} {\bibfnamefont {S.}~\bibnamefont {Carr}}, \bibinfo {author}
  {\bibfnamefont {H.~G.}\ \bibnamefont {Brown}}, \bibinfo {author}
  {\bibfnamefont {J.}~\bibnamefont {Ciston}}, \bibinfo {author} {\bibfnamefont
  {T.}~\bibnamefont {Taniguchi}}, \bibinfo {author} {\bibfnamefont
  {K.}~\bibnamefont {Watanabe}},\ and\ \bibinfo {author} {\bibfnamefont
  {D.~K.}\ \bibnamefont {Bediako}},\ }\bibfield  {title} {\bibinfo {title}
  {Strain fields in twisted bilayer graphene},\ }\href
  {https://doi.org/10.1038/s41563-021-00973-w} {\bibfield  {journal} {\bibinfo
  {journal} {Nature Materials}\ }\textbf {\bibinfo {volume} {20}},\ \bibinfo
  {pages} {956} (\bibinfo {year} {2021})}\BibitemShut {NoStop}%
\bibitem [{\citenamefont {Polshyn}\ \emph {et~al.}(2020)\citenamefont
  {Polshyn}, \citenamefont {Zhu}, \citenamefont {Kumar}, \citenamefont {Zhang},
  \citenamefont {Yang}, \citenamefont {Tschirhart}, \citenamefont {Serlin},
  \citenamefont {Watanabe}, \citenamefont {Taniguchi}, \citenamefont
  {MacDonald} \emph {et~al.}}]{polshyn2020}%
  \BibitemOpen
  \bibfield  {author} {\bibinfo {author} {\bibfnamefont {H.}~\bibnamefont
  {Polshyn}}, \bibinfo {author} {\bibfnamefont {J.}~\bibnamefont {Zhu}},
  \bibinfo {author} {\bibfnamefont {M.~A.}\ \bibnamefont {Kumar}}, \bibinfo
  {author} {\bibfnamefont {Y.}~\bibnamefont {Zhang}}, \bibinfo {author}
  {\bibfnamefont {F.}~\bibnamefont {Yang}}, \bibinfo {author} {\bibfnamefont
  {C.~L.}\ \bibnamefont {Tschirhart}}, \bibinfo {author} {\bibfnamefont
  {M.}~\bibnamefont {Serlin}}, \bibinfo {author} {\bibfnamefont
  {K.}~\bibnamefont {Watanabe}}, \bibinfo {author} {\bibfnamefont
  {T.}~\bibnamefont {Taniguchi}}, \bibinfo {author} {\bibfnamefont {A.~H.}\
  \bibnamefont {MacDonald}}, \emph {et~al.},\ }\bibfield  {title} {\bibinfo
  {title} {Electrical switching of magnetic order in an orbital chern
  insulator},\ }\href {https://doi.org/10.1038/s41586-020-2963-8} {\bibfield
  {journal} {\bibinfo  {journal} {Nature}\ }\textbf {\bibinfo {volume} {588}},\
  \bibinfo {pages} {66} (\bibinfo {year} {2020})}\BibitemShut {NoStop}%
\bibitem [{\citenamefont {Xu}\ \emph {et~al.}(2021)\citenamefont {Xu},
  \citenamefont {Al~Ezzi}, \citenamefont {Balakrishnan}, \citenamefont
  {Garcia-Ruiz}, \citenamefont {Tsim}, \citenamefont {Mullan}, \citenamefont
  {Barrier}, \citenamefont {Xin}, \citenamefont {Piot}, \citenamefont
  {Taniguchi}, \citenamefont {Watanabe}, \citenamefont {Carvalho},
  \citenamefont {Mishchenko}, \citenamefont {Geim}, \citenamefont {Fal'ko},
  \citenamefont {Adam}, \citenamefont {Neto}, \citenamefont {Novoselov},\ and\
  \citenamefont {Shi}}]{Xu2021}%
  \BibitemOpen
  \bibfield  {author} {\bibinfo {author} {\bibfnamefont {S.}~\bibnamefont
  {Xu}}, \bibinfo {author} {\bibfnamefont {M.~M.}\ \bibnamefont {Al~Ezzi}},
  \bibinfo {author} {\bibfnamefont {N.}~\bibnamefont {Balakrishnan}}, \bibinfo
  {author} {\bibfnamefont {A.}~\bibnamefont {Garcia-Ruiz}}, \bibinfo {author}
  {\bibfnamefont {B.}~\bibnamefont {Tsim}}, \bibinfo {author} {\bibfnamefont
  {C.}~\bibnamefont {Mullan}}, \bibinfo {author} {\bibfnamefont
  {J.}~\bibnamefont {Barrier}}, \bibinfo {author} {\bibfnamefont
  {N.}~\bibnamefont {Xin}}, \bibinfo {author} {\bibfnamefont {B.~A.}\
  \bibnamefont {Piot}}, \bibinfo {author} {\bibfnamefont {T.}~\bibnamefont
  {Taniguchi}}, \bibinfo {author} {\bibfnamefont {K.}~\bibnamefont {Watanabe}},
  \bibinfo {author} {\bibfnamefont {A.}~\bibnamefont {Carvalho}}, \bibinfo
  {author} {\bibfnamefont {A.}~\bibnamefont {Mishchenko}}, \bibinfo {author}
  {\bibfnamefont {A.~K.}\ \bibnamefont {Geim}}, \bibinfo {author}
  {\bibfnamefont {V.~I.}\ \bibnamefont {Fal'ko}}, \bibinfo {author}
  {\bibfnamefont {S.}~\bibnamefont {Adam}}, \bibinfo {author} {\bibfnamefont
  {A.~H.~C.}\ \bibnamefont {Neto}}, \bibinfo {author} {\bibfnamefont {K.~S.}\
  \bibnamefont {Novoselov}},\ and\ \bibinfo {author} {\bibfnamefont
  {Y.}~\bibnamefont {Shi}},\ }\bibfield  {title} {\bibinfo {title} {Tunable van
  hove singularities and correlated states in twisted monolayer--bilayer
  graphene},\ }\href {https://doi.org/10.1038/s41567-021-01172-9} {\bibfield
  {journal} {\bibinfo  {journal} {Nature Physics}\ }\textbf {\bibinfo {volume}
  {17}},\ \bibinfo {pages} {619} (\bibinfo {year} {2021})}\BibitemShut
  {NoStop}%
\bibitem [{\citenamefont {Chen}\ \emph {et~al.}(2019)\citenamefont {Chen},
  \citenamefont {Jiang}, \citenamefont {Wu}, \citenamefont {Lyu}, \citenamefont
  {Li}, \citenamefont {Chittari}, \citenamefont {Watanabe}, \citenamefont
  {Taniguchi}, \citenamefont {Shi}, \citenamefont {Jung}, \citenamefont
  {Zhang},\ and\ \citenamefont {Wang}}]{Chen2019}%
  \BibitemOpen
  \bibfield  {author} {\bibinfo {author} {\bibfnamefont {G.}~\bibnamefont
  {Chen}}, \bibinfo {author} {\bibfnamefont {L.}~\bibnamefont {Jiang}},
  \bibinfo {author} {\bibfnamefont {S.}~\bibnamefont {Wu}}, \bibinfo {author}
  {\bibfnamefont {B.}~\bibnamefont {Lyu}}, \bibinfo {author} {\bibfnamefont
  {H.}~\bibnamefont {Li}}, \bibinfo {author} {\bibfnamefont {B.~L.}\
  \bibnamefont {Chittari}}, \bibinfo {author} {\bibfnamefont {K.}~\bibnamefont
  {Watanabe}}, \bibinfo {author} {\bibfnamefont {T.}~\bibnamefont {Taniguchi}},
  \bibinfo {author} {\bibfnamefont {Z.}~\bibnamefont {Shi}}, \bibinfo {author}
  {\bibfnamefont {J.}~\bibnamefont {Jung}}, \bibinfo {author} {\bibfnamefont
  {Y.}~\bibnamefont {Zhang}},\ and\ \bibinfo {author} {\bibfnamefont
  {F.}~\bibnamefont {Wang}},\ }\bibfield  {title} {\bibinfo {title} {Evidence
  of a gate-tunable mott insulator in a trilayer graphene moir{\'e}
  superlattice},\ }\href {https://doi.org/10.1038/s41567-018-0387-2} {\bibfield
   {journal} {\bibinfo  {journal} {Nature Physics}\ }\textbf {\bibinfo {volume}
  {15}},\ \bibinfo {pages} {237} (\bibinfo {year} {2019})}\BibitemShut
  {NoStop}%
\bibitem [{\citenamefont {Shen}\ \emph {et~al.}(2020)\citenamefont {Shen},
  \citenamefont {Chu}, \citenamefont {Wu}, \citenamefont {Li}, \citenamefont
  {Wang}, \citenamefont {Zhao}, \citenamefont {Tang}, \citenamefont {Liu},
  \citenamefont {Tian}, \citenamefont {Watanabe} \emph {et~al.}}]{shen2020}%
  \BibitemOpen
  \bibfield  {author} {\bibinfo {author} {\bibfnamefont {C.}~\bibnamefont
  {Shen}}, \bibinfo {author} {\bibfnamefont {Y.}~\bibnamefont {Chu}}, \bibinfo
  {author} {\bibfnamefont {Q.}~\bibnamefont {Wu}}, \bibinfo {author}
  {\bibfnamefont {N.}~\bibnamefont {Li}}, \bibinfo {author} {\bibfnamefont
  {S.}~\bibnamefont {Wang}}, \bibinfo {author} {\bibfnamefont {Y.}~\bibnamefont
  {Zhao}}, \bibinfo {author} {\bibfnamefont {J.}~\bibnamefont {Tang}}, \bibinfo
  {author} {\bibfnamefont {J.}~\bibnamefont {Liu}}, \bibinfo {author}
  {\bibfnamefont {J.}~\bibnamefont {Tian}}, \bibinfo {author} {\bibfnamefont
  {K.}~\bibnamefont {Watanabe}}, \emph {et~al.},\ }\bibfield  {title} {\bibinfo
  {title} {Correlated states in twisted double bilayer graphene},\ }\href
  {https://doi.org/10.1038/s41567-020-0825-9} {\bibfield  {journal} {\bibinfo
  {journal} {Nature Physics}\ }\textbf {\bibinfo {volume} {16}},\ \bibinfo
  {pages} {520} (\bibinfo {year} {2020})}\BibitemShut {NoStop}%
\bibitem [{\citenamefont {Chen}\ \emph {et~al.}(2021)\citenamefont {Chen},
  \citenamefont {He}, \citenamefont {Zhang}, \citenamefont {Hsieh},
  \citenamefont {Fei}, \citenamefont {Watanabe}, \citenamefont {Taniguchi},
  \citenamefont {Cobden}, \citenamefont {Xu}, \citenamefont {Dean} \emph
  {et~al.}}]{chen2021}%
  \BibitemOpen
  \bibfield  {author} {\bibinfo {author} {\bibfnamefont {S.}~\bibnamefont
  {Chen}}, \bibinfo {author} {\bibfnamefont {M.}~\bibnamefont {He}}, \bibinfo
  {author} {\bibfnamefont {Y.-H.}\ \bibnamefont {Zhang}}, \bibinfo {author}
  {\bibfnamefont {V.}~\bibnamefont {Hsieh}}, \bibinfo {author} {\bibfnamefont
  {Z.}~\bibnamefont {Fei}}, \bibinfo {author} {\bibfnamefont {K.}~\bibnamefont
  {Watanabe}}, \bibinfo {author} {\bibfnamefont {T.}~\bibnamefont {Taniguchi}},
  \bibinfo {author} {\bibfnamefont {D.~H.}\ \bibnamefont {Cobden}}, \bibinfo
  {author} {\bibfnamefont {X.}~\bibnamefont {Xu}}, \bibinfo {author}
  {\bibfnamefont {C.~R.}\ \bibnamefont {Dean}}, \emph {et~al.},\ }\bibfield
  {title} {\bibinfo {title} {Electrically tunable correlated and topological
  states in twisted monolayer--bilayer graphene},\ }\href
  {https://doi.org/10.1038/s41567-020-01062-6} {\bibfield  {journal} {\bibinfo
  {journal} {Nature Physics}\ }\textbf {\bibinfo {volume} {17}},\ \bibinfo
  {pages} {374} (\bibinfo {year} {2021})}\BibitemShut {NoStop}%
\bibitem [{\citenamefont {Park}\ \emph {et~al.}(2021)\citenamefont {Park},
  \citenamefont {Cao}, \citenamefont {Watanabe}, \citenamefont {Taniguchi},\
  and\ \citenamefont {Jarillo-Herrero}}]{park2021}%
  \BibitemOpen
  \bibfield  {author} {\bibinfo {author} {\bibfnamefont {J.~M.}\ \bibnamefont
  {Park}}, \bibinfo {author} {\bibfnamefont {Y.}~\bibnamefont {Cao}}, \bibinfo
  {author} {\bibfnamefont {K.}~\bibnamefont {Watanabe}}, \bibinfo {author}
  {\bibfnamefont {T.}~\bibnamefont {Taniguchi}},\ and\ \bibinfo {author}
  {\bibfnamefont {P.}~\bibnamefont {Jarillo-Herrero}},\ }\bibfield  {title}
  {\bibinfo {title} {Tunable strongly coupled superconductivity in magic-angle
  twisted trilayer graphene},\ }\href
  {https://doi.org/doi.org/10.1038/s41586-021-03192-0} {\bibfield  {journal}
  {\bibinfo  {journal} {Nature}\ }\textbf {\bibinfo {volume} {590}},\ \bibinfo
  {pages} {249} (\bibinfo {year} {2021})}\BibitemShut {NoStop}%
\bibitem [{\citenamefont {Kunstmann}\ \emph {et~al.}(2018)\citenamefont
  {Kunstmann}, \citenamefont {Mooshammer}, \citenamefont {Nagler},
  \citenamefont {Chaves}, \citenamefont {Stein}, \citenamefont {Paradiso},
  \citenamefont {Plechinger}, \citenamefont {Strunk}, \citenamefont
  {Sch{\"u}ller}, \citenamefont {Seifert},\ and\ \citenamefont
  {et~al}}]{kunstmann2018momentum}%
  \BibitemOpen
  \bibfield  {author} {\bibinfo {author} {\bibfnamefont {J.}~\bibnamefont
  {Kunstmann}}, \bibinfo {author} {\bibfnamefont {F.}~\bibnamefont
  {Mooshammer}}, \bibinfo {author} {\bibfnamefont {P.}~\bibnamefont {Nagler}},
  \bibinfo {author} {\bibfnamefont {A.}~\bibnamefont {Chaves}}, \bibinfo
  {author} {\bibfnamefont {F.}~\bibnamefont {Stein}}, \bibinfo {author}
  {\bibfnamefont {N.}~\bibnamefont {Paradiso}}, \bibinfo {author}
  {\bibfnamefont {G.}~\bibnamefont {Plechinger}}, \bibinfo {author}
  {\bibfnamefont {C.}~\bibnamefont {Strunk}}, \bibinfo {author} {\bibfnamefont
  {C.}~\bibnamefont {Sch{\"u}ller}}, \bibinfo {author} {\bibfnamefont
  {G.}~\bibnamefont {Seifert}},\ and\ \bibinfo {author} {\bibnamefont
  {et~al}},\ }\bibfield  {title} {\bibinfo {title} {Momentum-space indirect
  interlayer excitons in transition-metal dichalcogenide van der {W}aals
  heterostructures},\ }\href {https://doi.org/10.1038/s41567-018-0123-y}
  {\bibfield  {journal} {\bibinfo  {journal} {Nature Physics}\ }\textbf
  {\bibinfo {volume} {14}},\ \bibinfo {pages} {801} (\bibinfo {year}
  {2018})}\BibitemShut {NoStop}%
\bibitem [{\citenamefont {Rivera}\ \emph {et~al.}(2018)\citenamefont {Rivera},
  \citenamefont {Yu}, \citenamefont {Seyler}, \citenamefont {Wilson},
  \citenamefont {Yao},\ and\ \citenamefont {Xu}}]{rivera2018interlayer}%
  \BibitemOpen
  \bibfield  {author} {\bibinfo {author} {\bibfnamefont {P.}~\bibnamefont
  {Rivera}}, \bibinfo {author} {\bibfnamefont {H.}~\bibnamefont {Yu}}, \bibinfo
  {author} {\bibfnamefont {K.~L.}\ \bibnamefont {Seyler}}, \bibinfo {author}
  {\bibfnamefont {N.~P.}\ \bibnamefont {Wilson}}, \bibinfo {author}
  {\bibfnamefont {W.}~\bibnamefont {Yao}},\ and\ \bibinfo {author}
  {\bibfnamefont {X.}~\bibnamefont {Xu}},\ }\bibfield  {title} {\bibinfo
  {title} {Interlayer valley excitons in heterobilayers of transition metal
  dichalcogenides},\ }\href {https://doi.org/10.1038/s41565-018-0193-0}
  {\bibfield  {journal} {\bibinfo  {journal} {Nature Nanotechnology}\ }\textbf
  {\bibinfo {volume} {13}},\ \bibinfo {pages} {1004} (\bibinfo {year}
  {2018})}\BibitemShut {NoStop}%
\bibitem [{\citenamefont {Nayak}\ \emph {et~al.}(2017)\citenamefont {Nayak},
  \citenamefont {Horbatenko}, \citenamefont {Ahn}, \citenamefont {Kim},
  \citenamefont {Lee}, \citenamefont {Ma}, \citenamefont {Jang}, \citenamefont
  {Lim}, \citenamefont {Kim}, \citenamefont {Ryu},\ and\ \citenamefont
  {et~al}}]{nayak2017probing}%
  \BibitemOpen
  \bibfield  {author} {\bibinfo {author} {\bibfnamefont {P.~K.}\ \bibnamefont
  {Nayak}}, \bibinfo {author} {\bibfnamefont {Y.}~\bibnamefont {Horbatenko}},
  \bibinfo {author} {\bibfnamefont {S.}~\bibnamefont {Ahn}}, \bibinfo {author}
  {\bibfnamefont {G.}~\bibnamefont {Kim}}, \bibinfo {author} {\bibfnamefont
  {J.-U.}\ \bibnamefont {Lee}}, \bibinfo {author} {\bibfnamefont {K.~Y.}\
  \bibnamefont {Ma}}, \bibinfo {author} {\bibfnamefont {A.-R.}\ \bibnamefont
  {Jang}}, \bibinfo {author} {\bibfnamefont {H.}~\bibnamefont {Lim}}, \bibinfo
  {author} {\bibfnamefont {D.}~\bibnamefont {Kim}}, \bibinfo {author}
  {\bibfnamefont {S.}~\bibnamefont {Ryu}},\ and\ \bibinfo {author}
  {\bibnamefont {et~al}},\ }\bibfield  {title} {\bibinfo {title} {Probing
  evolution of twist-angle-dependent interlayer excitons in
  {MoSe}$_2$/{WSe}$_2$ van der {W}aals heterostructures},\ }\href
  {https://doi.org/10.1021/acsnano.7b00640} {\bibfield  {journal} {\bibinfo
  {journal} {ACS Nano}\ }\textbf {\bibinfo {volume} {11}},\ \bibinfo {pages}
  {4041} (\bibinfo {year} {2017})}\BibitemShut {NoStop}%
\bibitem [{\citenamefont {Alexeev}\ \emph {et~al.}(2019)\citenamefont
  {Alexeev}, \citenamefont {Ruiz-Tijerina}, \citenamefont {Danovich},
  \citenamefont {Hamer}, \citenamefont {Terry}, \citenamefont {Nayak},
  \citenamefont {Ahn}, \citenamefont {Pak}, \citenamefont {Lee}, \citenamefont
  {Sohn},\ and\ \citenamefont {et~al}}]{alexeev2019}%
  \BibitemOpen
  \bibfield  {author} {\bibinfo {author} {\bibfnamefont {E.~M.}\ \bibnamefont
  {Alexeev}}, \bibinfo {author} {\bibfnamefont {D.~A.}\ \bibnamefont
  {Ruiz-Tijerina}}, \bibinfo {author} {\bibfnamefont {M.}~\bibnamefont
  {Danovich}}, \bibinfo {author} {\bibfnamefont {M.~J.}\ \bibnamefont {Hamer}},
  \bibinfo {author} {\bibfnamefont {D.~J.}\ \bibnamefont {Terry}}, \bibinfo
  {author} {\bibfnamefont {P.~K.}\ \bibnamefont {Nayak}}, \bibinfo {author}
  {\bibfnamefont {S.}~\bibnamefont {Ahn}}, \bibinfo {author} {\bibfnamefont
  {S.}~\bibnamefont {Pak}}, \bibinfo {author} {\bibfnamefont {J.}~\bibnamefont
  {Lee}}, \bibinfo {author} {\bibfnamefont {J.~I.}\ \bibnamefont {Sohn}},\ and\
  \bibinfo {author} {\bibnamefont {et~al}},\ }\bibfield  {title} {\bibinfo
  {title} {Resonantly hybridized excitons in moir{\'e} superlattices in van der
  {W}aals heterostructures},\ }\href
  {https://doi.org/10.1038/s41586-019-0986-9} {\bibfield  {journal} {\bibinfo
  {journal} {Nature}\ }\textbf {\bibinfo {volume} {567}},\ \bibinfo {pages}
  {81} (\bibinfo {year} {2019})}\BibitemShut {NoStop}%
\bibitem [{\citenamefont {McGilly}\ \emph {et~al.}(2020)\citenamefont
  {McGilly}, \citenamefont {Kerelsky}, \citenamefont {Finney}, \citenamefont
  {Shapovalov}, \citenamefont {Shih}, \citenamefont {Ghiotto}, \citenamefont
  {Zeng}, \citenamefont {Moore}, \citenamefont {Wu}, \citenamefont {Bai},
  \citenamefont {Watanabe}, \citenamefont {Taniguchi}, \citenamefont {Stengel},
  \citenamefont {Zhou}, \citenamefont {Hone}, \citenamefont {Zhu},
  \citenamefont {Basov}, \citenamefont {Dean}, \citenamefont {Dreyer},\ and\
  \citenamefont {Pasupathy}}]{McGilly2020}%
  \BibitemOpen
  \bibfield  {author} {\bibinfo {author} {\bibfnamefont {L.~J.}\ \bibnamefont
  {McGilly}}, \bibinfo {author} {\bibfnamefont {A.}~\bibnamefont {Kerelsky}},
  \bibinfo {author} {\bibfnamefont {N.~R.}\ \bibnamefont {Finney}}, \bibinfo
  {author} {\bibfnamefont {K.}~\bibnamefont {Shapovalov}}, \bibinfo {author}
  {\bibfnamefont {E.-M.}\ \bibnamefont {Shih}}, \bibinfo {author}
  {\bibfnamefont {A.}~\bibnamefont {Ghiotto}}, \bibinfo {author} {\bibfnamefont
  {Y.}~\bibnamefont {Zeng}}, \bibinfo {author} {\bibfnamefont {S.~L.}\
  \bibnamefont {Moore}}, \bibinfo {author} {\bibfnamefont {W.}~\bibnamefont
  {Wu}}, \bibinfo {author} {\bibfnamefont {Y.}~\bibnamefont {Bai}}, \bibinfo
  {author} {\bibfnamefont {K.}~\bibnamefont {Watanabe}}, \bibinfo {author}
  {\bibfnamefont {T.}~\bibnamefont {Taniguchi}}, \bibinfo {author}
  {\bibfnamefont {M.}~\bibnamefont {Stengel}}, \bibinfo {author} {\bibfnamefont
  {L.}~\bibnamefont {Zhou}}, \bibinfo {author} {\bibfnamefont {J.}~\bibnamefont
  {Hone}}, \bibinfo {author} {\bibfnamefont {X.}~\bibnamefont {Zhu}}, \bibinfo
  {author} {\bibfnamefont {D.~N.}\ \bibnamefont {Basov}}, \bibinfo {author}
  {\bibfnamefont {C.}~\bibnamefont {Dean}}, \bibinfo {author} {\bibfnamefont
  {C.~E.}\ \bibnamefont {Dreyer}},\ and\ \bibinfo {author} {\bibfnamefont
  {A.~N.}\ \bibnamefont {Pasupathy}},\ }\bibfield  {title} {\bibinfo {title}
  {Visualization of moir\'e superlattices},\ }\href
  {https://doi.org/10.1038/s41565-020-0708-3} {\bibfield  {journal} {\bibinfo
  {journal} {Nature Nanotechnology}\ }\textbf {\bibinfo {volume} {15}},\
  \bibinfo {pages} {580} (\bibinfo {year} {2020})}\BibitemShut {NoStop}%
\bibitem [{\citenamefont {Zhang}\ \emph {et~al.}(2020)\citenamefont {Zhang},
  \citenamefont {Wang}, \citenamefont {Watanabe}, \citenamefont {Taniguchi},
  \citenamefont {Ueno}, \citenamefont {Tutuc},\ and\ \citenamefont
  {LeRoy}}]{Zhang2020}%
  \BibitemOpen
  \bibfield  {author} {\bibinfo {author} {\bibfnamefont {Z.}~\bibnamefont
  {Zhang}}, \bibinfo {author} {\bibfnamefont {Y.}~\bibnamefont {Wang}},
  \bibinfo {author} {\bibfnamefont {K.}~\bibnamefont {Watanabe}}, \bibinfo
  {author} {\bibfnamefont {T.}~\bibnamefont {Taniguchi}}, \bibinfo {author}
  {\bibfnamefont {K.}~\bibnamefont {Ueno}}, \bibinfo {author} {\bibfnamefont
  {E.}~\bibnamefont {Tutuc}},\ and\ \bibinfo {author} {\bibfnamefont {B.~J.}\
  \bibnamefont {LeRoy}},\ }\bibfield  {title} {\bibinfo {title} {Flat bands in
  twisted bilayer transition metal dichalcogenides},\ }\href
  {https://doi.org/10.1038/s41567-020-0958-x} {\bibfield  {journal} {\bibinfo
  {journal} {Nature Physics}\ }\textbf {\bibinfo {volume} {16}},\ \bibinfo
  {pages} {1093} (\bibinfo {year} {2020})}\BibitemShut {NoStop}%
\bibitem [{\citenamefont {Wang}\ \emph {et~al.}(2020)\citenamefont {Wang},
  \citenamefont {Shih}, \citenamefont {Ghiotto}, \citenamefont {Xian},
  \citenamefont {Rhodes}, \citenamefont {Tan}, \citenamefont {Claassen},
  \citenamefont {Kennes}, \citenamefont {Bai}, \citenamefont {Kim},
  \citenamefont {Watanabe}, \citenamefont {Taniguchi}, \citenamefont {Zhu},
  \citenamefont {Hone}, \citenamefont {Rubio}, \citenamefont {Pasupathy},\ and\
  \citenamefont {Dean}}]{Wang2020}%
  \BibitemOpen
  \bibfield  {author} {\bibinfo {author} {\bibfnamefont {L.}~\bibnamefont
  {Wang}}, \bibinfo {author} {\bibfnamefont {E.-M.}\ \bibnamefont {Shih}},
  \bibinfo {author} {\bibfnamefont {A.}~\bibnamefont {Ghiotto}}, \bibinfo
  {author} {\bibfnamefont {L.}~\bibnamefont {Xian}}, \bibinfo {author}
  {\bibfnamefont {D.~A.}\ \bibnamefont {Rhodes}}, \bibinfo {author}
  {\bibfnamefont {C.}~\bibnamefont {Tan}}, \bibinfo {author} {\bibfnamefont
  {M.}~\bibnamefont {Claassen}}, \bibinfo {author} {\bibfnamefont {D.~M.}\
  \bibnamefont {Kennes}}, \bibinfo {author} {\bibfnamefont {Y.}~\bibnamefont
  {Bai}}, \bibinfo {author} {\bibfnamefont {B.}~\bibnamefont {Kim}}, \bibinfo
  {author} {\bibfnamefont {K.}~\bibnamefont {Watanabe}}, \bibinfo {author}
  {\bibfnamefont {T.}~\bibnamefont {Taniguchi}}, \bibinfo {author}
  {\bibfnamefont {X.}~\bibnamefont {Zhu}}, \bibinfo {author} {\bibfnamefont
  {J.}~\bibnamefont {Hone}}, \bibinfo {author} {\bibfnamefont {A.}~\bibnamefont
  {Rubio}}, \bibinfo {author} {\bibfnamefont {A.~N.}\ \bibnamefont
  {Pasupathy}},\ and\ \bibinfo {author} {\bibfnamefont {C.~R.}\ \bibnamefont
  {Dean}},\ }\bibfield  {title} {\bibinfo {title} {Correlated electronic phases
  in twisted bilayer transition metal dichalcogenides},\ }\href
  {https://doi.org/10.1038/s41563-020-0708-6} {\bibfield  {journal} {\bibinfo
  {journal} {Nature Materials}\ }\textbf {\bibinfo {volume} {19}},\ \bibinfo
  {pages} {861} (\bibinfo {year} {2020})}\BibitemShut {NoStop}%
\bibitem [{\citenamefont {Alden}\ \emph {et~al.}(2013)\citenamefont {Alden},
  \citenamefont {Tsen}, \citenamefont {Huang}, \citenamefont {Hovden},
  \citenamefont {Brown}, \citenamefont {Park}, \citenamefont {Muller},\ and\
  \citenamefont {McEuen}}]{AldenPNAS}%
  \BibitemOpen
  \bibfield  {author} {\bibinfo {author} {\bibfnamefont {J.~S.}\ \bibnamefont
  {Alden}}, \bibinfo {author} {\bibfnamefont {A.~W.}\ \bibnamefont {Tsen}},
  \bibinfo {author} {\bibfnamefont {P.~Y.}\ \bibnamefont {Huang}}, \bibinfo
  {author} {\bibfnamefont {R.}~\bibnamefont {Hovden}}, \bibinfo {author}
  {\bibfnamefont {L.}~\bibnamefont {Brown}}, \bibinfo {author} {\bibfnamefont
  {J.}~\bibnamefont {Park}}, \bibinfo {author} {\bibfnamefont {D.~A.}\
  \bibnamefont {Muller}},\ and\ \bibinfo {author} {\bibfnamefont {P.~L.}\
  \bibnamefont {McEuen}},\ }\bibfield  {title} {\bibinfo {title} {Strain
  solitons and topological defects in bilayer graphene},\ }\href
  {https://doi.org/10.1073/pnas.1309394110} {\bibfield  {journal} {\bibinfo
  {journal} {PNAS}\ }\textbf {\bibinfo {volume} {110}},\ \bibinfo {pages}
  {11256} (\bibinfo {year} {2013})}\BibitemShut {NoStop}%
\bibitem [{\citenamefont {Yoo}\ \emph {et~al.}(2019)\citenamefont {Yoo},
  \citenamefont {Engelke}, \citenamefont {Carr}, \citenamefont {Fang},
  \citenamefont {Zhang}, \citenamefont {Cazeaux}, \citenamefont {Sung},
  \citenamefont {Hovden}, \citenamefont {Tsen}, \citenamefont {Taniguchi},\
  and\ \citenamefont {et~al}}]{yoo2019atomic}%
  \BibitemOpen
  \bibfield  {author} {\bibinfo {author} {\bibfnamefont {H.}~\bibnamefont
  {Yoo}}, \bibinfo {author} {\bibfnamefont {R.}~\bibnamefont {Engelke}},
  \bibinfo {author} {\bibfnamefont {S.}~\bibnamefont {Carr}}, \bibinfo {author}
  {\bibfnamefont {S.}~\bibnamefont {Fang}}, \bibinfo {author} {\bibfnamefont
  {K.}~\bibnamefont {Zhang}}, \bibinfo {author} {\bibfnamefont
  {P.}~\bibnamefont {Cazeaux}}, \bibinfo {author} {\bibfnamefont {S.~H.}\
  \bibnamefont {Sung}}, \bibinfo {author} {\bibfnamefont {R.}~\bibnamefont
  {Hovden}}, \bibinfo {author} {\bibfnamefont {A.~W.}\ \bibnamefont {Tsen}},
  \bibinfo {author} {\bibfnamefont {T.}~\bibnamefont {Taniguchi}},\ and\
  \bibinfo {author} {\bibnamefont {et~al}},\ }\bibfield  {title} {\bibinfo
  {title} {Atomic and electronic reconstruction at the van der {W}aals
  interface in twisted bilayer graphene},\ }\href
  {https://doi.org/10.1038/s41563-019-0346-z} {\bibfield  {journal} {\bibinfo
  {journal} {Nature materials}\ }\textbf {\bibinfo {volume} {18}},\ \bibinfo
  {pages} {448} (\bibinfo {year} {2019})}\BibitemShut {NoStop}%
\bibitem [{\citenamefont {Rosenberger}\ \emph {et~al.}(2020)\citenamefont
  {Rosenberger}, \citenamefont {Chuang}, \citenamefont {Phillips},
  \citenamefont {Oleshko}, \citenamefont {McCreary}, \citenamefont {Sivaram},
  \citenamefont {Hellberg},\ and\ \citenamefont {Jonker}}]{rosenberger2020}%
  \BibitemOpen
  \bibfield  {author} {\bibinfo {author} {\bibfnamefont {M.~R.}\ \bibnamefont
  {Rosenberger}}, \bibinfo {author} {\bibfnamefont {H.-J.}\ \bibnamefont
  {Chuang}}, \bibinfo {author} {\bibfnamefont {M.}~\bibnamefont {Phillips}},
  \bibinfo {author} {\bibfnamefont {V.~P.}\ \bibnamefont {Oleshko}}, \bibinfo
  {author} {\bibfnamefont {K.~M.}\ \bibnamefont {McCreary}}, \bibinfo {author}
  {\bibfnamefont {S.~V.}\ \bibnamefont {Sivaram}}, \bibinfo {author}
  {\bibfnamefont {C.~S.}\ \bibnamefont {Hellberg}},\ and\ \bibinfo {author}
  {\bibfnamefont {B.~T.}\ \bibnamefont {Jonker}},\ }\bibfield  {title}
  {\bibinfo {title} {Twist angle-dependent atomic reconstruction and moir{\'e}
  patterns in transition metal dichalcogenide heterostructures},\ }\href
  {https://doi.org/10.1021/acsnano.0c00088} {\bibfield  {journal} {\bibinfo
  {journal} {ACS Nano}\ }\textbf {\bibinfo {volume} {14}},\ \bibinfo {pages}
  {4550} (\bibinfo {year} {2020})}\BibitemShut {NoStop}%
\bibitem [{\citenamefont {Weston}\ \emph {et~al.}(2020)\citenamefont {Weston},
  \citenamefont {Zou}, \citenamefont {Enaldiev}, \citenamefont {Summerfield},
  \citenamefont {Clark}, \citenamefont {Z{\'o}lyomi}, \citenamefont {Graham},
  \citenamefont {Yelgel}, \citenamefont {Magorrian}, \citenamefont {Zhou},
  \citenamefont {Zultak}, \citenamefont {Hopkinson}, \citenamefont {Barinov},
  \citenamefont {Bointon}, \citenamefont {Kretinin}, \citenamefont {Wilson},
  \citenamefont {Beton}, \citenamefont {Fal'ko}, \citenamefont {Haigh},\ and\
  \citenamefont {Gorbachev}}]{Weston2020}%
  \BibitemOpen
  \bibfield  {author} {\bibinfo {author} {\bibfnamefont {A.}~\bibnamefont
  {Weston}}, \bibinfo {author} {\bibfnamefont {Y.}~\bibnamefont {Zou}},
  \bibinfo {author} {\bibfnamefont {V.}~\bibnamefont {Enaldiev}}, \bibinfo
  {author} {\bibfnamefont {A.}~\bibnamefont {Summerfield}}, \bibinfo {author}
  {\bibfnamefont {N.}~\bibnamefont {Clark}}, \bibinfo {author} {\bibfnamefont
  {V.}~\bibnamefont {Z{\'o}lyomi}}, \bibinfo {author} {\bibfnamefont
  {A.}~\bibnamefont {Graham}}, \bibinfo {author} {\bibfnamefont
  {C.}~\bibnamefont {Yelgel}}, \bibinfo {author} {\bibfnamefont
  {S.}~\bibnamefont {Magorrian}}, \bibinfo {author} {\bibfnamefont
  {M.}~\bibnamefont {Zhou}}, \bibinfo {author} {\bibfnamefont {J.}~\bibnamefont
  {Zultak}}, \bibinfo {author} {\bibfnamefont {D.}~\bibnamefont {Hopkinson}},
  \bibinfo {author} {\bibfnamefont {A.}~\bibnamefont {Barinov}}, \bibinfo
  {author} {\bibfnamefont {T.~H.}\ \bibnamefont {Bointon}}, \bibinfo {author}
  {\bibfnamefont {A.}~\bibnamefont {Kretinin}}, \bibinfo {author}
  {\bibfnamefont {N.~R.}\ \bibnamefont {Wilson}}, \bibinfo {author}
  {\bibfnamefont {P.~H.}\ \bibnamefont {Beton}}, \bibinfo {author}
  {\bibfnamefont {V.~I.}\ \bibnamefont {Fal'ko}}, \bibinfo {author}
  {\bibfnamefont {S.~J.}\ \bibnamefont {Haigh}},\ and\ \bibinfo {author}
  {\bibfnamefont {R.}~\bibnamefont {Gorbachev}},\ }\bibfield  {title} {\bibinfo
  {title} {Atomic reconstruction in twisted bilayers of transition metal
  dichalcogenides},\ }\href {https://doi.org/10.1038/s41565-020-0682-9}
  {\bibfield  {journal} {\bibinfo  {journal} {Nature Nanotechnology}\ }\textbf
  {\bibinfo {volume} {15}},\ \bibinfo {pages} {592} (\bibinfo {year}
  {2020})}\BibitemShut {NoStop}%
\bibitem [{\citenamefont {Sung}\ \emph {et~al.}(2020)\citenamefont {Sung},
  \citenamefont {Zhou}, \citenamefont {Scuri}, \citenamefont {Z{\'o}lyomi},
  \citenamefont {Andersen}, \citenamefont {Yoo}, \citenamefont {Wild},
  \citenamefont {Joe}, \citenamefont {Gelly}, \citenamefont {Heo},
  \citenamefont {Magorrian}, \citenamefont {B{\'e}rub{\'e}}, \citenamefont
  {Valdivia}, \citenamefont {Taniguchi}, \citenamefont {Watanabe},
  \citenamefont {Lukin}, \citenamefont {Kim}, \citenamefont {Fal'ko},\ and\
  \citenamefont {Park}}]{Sung2020}%
  \BibitemOpen
  \bibfield  {author} {\bibinfo {author} {\bibfnamefont {J.}~\bibnamefont
  {Sung}}, \bibinfo {author} {\bibfnamefont {Y.}~\bibnamefont {Zhou}}, \bibinfo
  {author} {\bibfnamefont {G.}~\bibnamefont {Scuri}}, \bibinfo {author}
  {\bibfnamefont {V.}~\bibnamefont {Z{\'o}lyomi}}, \bibinfo {author}
  {\bibfnamefont {T.~I.}\ \bibnamefont {Andersen}}, \bibinfo {author}
  {\bibfnamefont {H.}~\bibnamefont {Yoo}}, \bibinfo {author} {\bibfnamefont
  {D.~S.}\ \bibnamefont {Wild}}, \bibinfo {author} {\bibfnamefont {A.~Y.}\
  \bibnamefont {Joe}}, \bibinfo {author} {\bibfnamefont {R.~J.}\ \bibnamefont
  {Gelly}}, \bibinfo {author} {\bibfnamefont {H.}~\bibnamefont {Heo}}, \bibinfo
  {author} {\bibfnamefont {S.~J.}\ \bibnamefont {Magorrian}}, \bibinfo {author}
  {\bibfnamefont {D.}~\bibnamefont {B{\'e}rub{\'e}}}, \bibinfo {author}
  {\bibfnamefont {A.~M.~M.}\ \bibnamefont {Valdivia}}, \bibinfo {author}
  {\bibfnamefont {T.}~\bibnamefont {Taniguchi}}, \bibinfo {author}
  {\bibfnamefont {K.}~\bibnamefont {Watanabe}}, \bibinfo {author}
  {\bibfnamefont {M.~D.}\ \bibnamefont {Lukin}}, \bibinfo {author}
  {\bibfnamefont {P.}~\bibnamefont {Kim}}, \bibinfo {author} {\bibfnamefont
  {V.~I.}\ \bibnamefont {Fal'ko}},\ and\ \bibinfo {author} {\bibfnamefont
  {H.}~\bibnamefont {Park}},\ }\bibfield  {title} {\bibinfo {title} {Broken
  mirror symmetry in excitonic response of reconstructed domains in twisted
  {MoSe}$_2$/{MoSe}$_2$ bilayers},\ }\href
  {https://doi.org/10.1038/s41565-020-0728-z} {\bibfield  {journal} {\bibinfo
  {journal} {Nature Nanotechnology}\ }\textbf {\bibinfo {volume} {15}},\
  \bibinfo {pages} {750} (\bibinfo {year} {2020})}\BibitemShut {NoStop}%
\bibitem [{\citenamefont {Woods}\ \emph {et~al.}(2021)\citenamefont {Woods},
  \citenamefont {Ares}, \citenamefont {Nevison-Andrews}, \citenamefont
  {Holwill}, \citenamefont {Fabregas}, \citenamefont {Guinea}, \citenamefont
  {Geim}, \citenamefont {Novoselov}, \citenamefont {Walet},\ and\ \citenamefont
  {Fumagalli}}]{woods2021}%
  \BibitemOpen
  \bibfield  {author} {\bibinfo {author} {\bibfnamefont {C.}~\bibnamefont
  {Woods}}, \bibinfo {author} {\bibfnamefont {P.}~\bibnamefont {Ares}},
  \bibinfo {author} {\bibfnamefont {H.}~\bibnamefont {Nevison-Andrews}},
  \bibinfo {author} {\bibfnamefont {M.}~\bibnamefont {Holwill}}, \bibinfo
  {author} {\bibfnamefont {R.}~\bibnamefont {Fabregas}}, \bibinfo {author}
  {\bibfnamefont {F.}~\bibnamefont {Guinea}}, \bibinfo {author} {\bibfnamefont
  {A.}~\bibnamefont {Geim}}, \bibinfo {author} {\bibfnamefont {K.}~\bibnamefont
  {Novoselov}}, \bibinfo {author} {\bibfnamefont {N.}~\bibnamefont {Walet}},\
  and\ \bibinfo {author} {\bibfnamefont {L.}~\bibnamefont {Fumagalli}},\
  }\bibfield  {title} {\bibinfo {title} {Charge-polarized interfacial
  superlattices in marginally twisted hexagonal boron nitride},\ }\href
  {https://doi.org/10.1038/s41467-020-20667-2} {\bibfield  {journal} {\bibinfo
  {journal} {Nature communications}\ }\textbf {\bibinfo {volume} {12}},\
  \bibinfo {pages} {1} (\bibinfo {year} {2021})}\BibitemShut {NoStop}%
\bibitem [{\citenamefont {Yasuda}\ \emph {et~al.}(2021)\citenamefont {Yasuda},
  \citenamefont {Wang}, \citenamefont {Watanabe}, \citenamefont {Taniguchi},\
  and\ \citenamefont {Jarillo-Herrero}}]{yasuda2021}%
  \BibitemOpen
  \bibfield  {author} {\bibinfo {author} {\bibfnamefont {K.}~\bibnamefont
  {Yasuda}}, \bibinfo {author} {\bibfnamefont {X.}~\bibnamefont {Wang}},
  \bibinfo {author} {\bibfnamefont {K.}~\bibnamefont {Watanabe}}, \bibinfo
  {author} {\bibfnamefont {T.}~\bibnamefont {Taniguchi}},\ and\ \bibinfo
  {author} {\bibfnamefont {P.}~\bibnamefont {Jarillo-Herrero}},\ }\bibfield
  {title} {\bibinfo {title} {Stacking-engineered ferroelectricity in bilayer
  boron nitride},\ }\href {https://doi.org/10.1126/science.abd3230} {\bibfield
  {journal} {\bibinfo  {journal} {Science}\ }\textbf {\bibinfo {volume}
  {372}},\ \bibinfo {pages} {1458} (\bibinfo {year} {2021})}\BibitemShut
  {NoStop}%
\bibitem [{\citenamefont {Stern}\ \emph {et~al.}(2021)\citenamefont {Stern},
  \citenamefont {Waschitz}, \citenamefont {Cao}, \citenamefont {Nevo},
  \citenamefont {Watanabe}, \citenamefont {Taniguchi}, \citenamefont {Sela},
  \citenamefont {Urbakh}, \citenamefont {Hod},\ and\ \citenamefont
  {Shalom}}]{stern2021}%
  \BibitemOpen
  \bibfield  {author} {\bibinfo {author} {\bibfnamefont {M.~V.}\ \bibnamefont
  {Stern}}, \bibinfo {author} {\bibfnamefont {Y.}~\bibnamefont {Waschitz}},
  \bibinfo {author} {\bibfnamefont {W.}~\bibnamefont {Cao}}, \bibinfo {author}
  {\bibfnamefont {I.}~\bibnamefont {Nevo}}, \bibinfo {author} {\bibfnamefont
  {K.}~\bibnamefont {Watanabe}}, \bibinfo {author} {\bibfnamefont
  {T.}~\bibnamefont {Taniguchi}}, \bibinfo {author} {\bibfnamefont
  {E.}~\bibnamefont {Sela}}, \bibinfo {author} {\bibfnamefont {M.}~\bibnamefont
  {Urbakh}}, \bibinfo {author} {\bibfnamefont {O.}~\bibnamefont {Hod}},\ and\
  \bibinfo {author} {\bibfnamefont {M.~B.}\ \bibnamefont {Shalom}},\ }\bibfield
   {title} {\bibinfo {title} {Interfacial ferroelectricity by van der {W}aals
  sliding},\ }\href {https://doi.org/10.1126/science.abe8177} {\bibfield
  {journal} {\bibinfo  {journal} {Science}\ }\textbf {\bibinfo {volume}
  {372}},\ \bibinfo {pages} {1462} (\bibinfo {year} {2021})}\BibitemShut
  {NoStop}%
\bibitem [{\citenamefont {Weston}\ \emph {et~al.}(2021)\citenamefont {Weston},
  \citenamefont {Castanon}, \citenamefont {Enaldiev}, \citenamefont {Ferreira},
  \citenamefont {Bhattacharjee}, \citenamefont {Xu}, \citenamefont
  {Corte-Leon}, \citenamefont {Wu}, \citenamefont {Clark}, \citenamefont
  {Summerfield} \emph {et~al.}}]{weston2021}%
  \BibitemOpen
  \bibfield  {author} {\bibinfo {author} {\bibfnamefont {A.}~\bibnamefont
  {Weston}}, \bibinfo {author} {\bibfnamefont {E.~G.}\ \bibnamefont
  {Castanon}}, \bibinfo {author} {\bibfnamefont {V.}~\bibnamefont {Enaldiev}},
  \bibinfo {author} {\bibfnamefont {F.}~\bibnamefont {Ferreira}}, \bibinfo
  {author} {\bibfnamefont {S.}~\bibnamefont {Bhattacharjee}}, \bibinfo {author}
  {\bibfnamefont {S.}~\bibnamefont {Xu}}, \bibinfo {author} {\bibfnamefont
  {H.}~\bibnamefont {Corte-Leon}}, \bibinfo {author} {\bibfnamefont
  {Z.}~\bibnamefont {Wu}}, \bibinfo {author} {\bibfnamefont {N.}~\bibnamefont
  {Clark}}, \bibinfo {author} {\bibfnamefont {A.}~\bibnamefont {Summerfield}},
  \emph {et~al.},\ }\bibfield  {title} {\bibinfo {title} {Interfacial
  ferroelectricity in marginally twisted 2{D} semiconductors},\ }\href
  {https://arxiv.org/abs/2108.06489} {\bibfield  {journal} {\bibinfo  {journal}
  {arXiv:2108.06489}\ } (\bibinfo {year} {2021})}\BibitemShut {NoStop}%
\bibitem [{\citenamefont {Wang}\ \emph {et~al.}(2021)\citenamefont {Wang},
  \citenamefont {Yasuda}, \citenamefont {Zhang}, \citenamefont {Liu},
  \citenamefont {Watanabe}, \citenamefont {Taniguchi}, \citenamefont {Hone},
  \citenamefont {Fu},\ and\ \citenamefont {Jarillo-Herrero}}]{wang2021}%
  \BibitemOpen
  \bibfield  {author} {\bibinfo {author} {\bibfnamefont {X.}~\bibnamefont
  {Wang}}, \bibinfo {author} {\bibfnamefont {K.}~\bibnamefont {Yasuda}},
  \bibinfo {author} {\bibfnamefont {Y.}~\bibnamefont {Zhang}}, \bibinfo
  {author} {\bibfnamefont {S.}~\bibnamefont {Liu}}, \bibinfo {author}
  {\bibfnamefont {K.}~\bibnamefont {Watanabe}}, \bibinfo {author}
  {\bibfnamefont {T.}~\bibnamefont {Taniguchi}}, \bibinfo {author}
  {\bibfnamefont {J.}~\bibnamefont {Hone}}, \bibinfo {author} {\bibfnamefont
  {L.}~\bibnamefont {Fu}},\ and\ \bibinfo {author} {\bibfnamefont
  {P.}~\bibnamefont {Jarillo-Herrero}},\ }\bibfield  {title} {\bibinfo {title}
  {Interfacial ferroelectricity in rhombohedral-stacked bilayer transition
  metal dichalcogenides},\ }\href {https://arxiv.org/abs/2108.07659} {\bibfield
   {journal} {\bibinfo  {journal} {arXiv:2108.07659}\ } (\bibinfo {year}
  {2021})}\BibitemShut {NoStop}%
\bibitem [{\citenamefont {Li}\ and\ \citenamefont {Wu}(2017)}]{li2017}%
  \BibitemOpen
  \bibfield  {author} {\bibinfo {author} {\bibfnamefont {L.}~\bibnamefont
  {Li}}\ and\ \bibinfo {author} {\bibfnamefont {M.}~\bibnamefont {Wu}},\
  }\bibfield  {title} {\bibinfo {title} {Binary compound bilayer and multilayer
  with vertical polarizations: two-dimensional ferroelectrics, multiferroics,
  and nanogenerators},\ }\href {https://doi.org/10.1021/acsnano.7b02756}
  {\bibfield  {journal} {\bibinfo  {journal} {ACS Nano}\ }\textbf {\bibinfo
  {volume} {11}},\ \bibinfo {pages} {6382} (\bibinfo {year}
  {2017})}\BibitemShut {NoStop}%
\bibitem [{\citenamefont {Tong}\ \emph {et~al.}(2020)\citenamefont {Tong},
  \citenamefont {Chen}, \citenamefont {Xiao}, \citenamefont {Yu},\ and\
  \citenamefont {Yao}}]{WangYao2020}%
  \BibitemOpen
  \bibfield  {author} {\bibinfo {author} {\bibfnamefont {Q.}~\bibnamefont
  {Tong}}, \bibinfo {author} {\bibfnamefont {M.}~\bibnamefont {Chen}}, \bibinfo
  {author} {\bibfnamefont {F.}~\bibnamefont {Xiao}}, \bibinfo {author}
  {\bibfnamefont {H.}~\bibnamefont {Yu}},\ and\ \bibinfo {author}
  {\bibfnamefont {W.}~\bibnamefont {Yao}},\ }\bibfield  {title} {\bibinfo
  {title} {Interferences of electrostatic moir{\'e} potentials and bichromatic
  superlattices of electrons and excitons in transition metal
  dichalcogenides},\ }\href {https://doi.org/10.1088/2053-1583/abd006}
  {\bibfield  {journal} {\bibinfo  {journal} {2D Materials}\ }\textbf {\bibinfo
  {volume} {8}},\ \bibinfo {pages} {025007} (\bibinfo {year}
  {2020})}\BibitemShut {NoStop}%
\bibitem [{\citenamefont {Ferreira}\ \emph
  {et~al.}(2021{\natexlab{a}})\citenamefont {Ferreira}, \citenamefont
  {Enaldiev}, \citenamefont {Fal'ko},\ and\ \citenamefont
  {Magorrian}}]{Ferreira2021}%
  \BibitemOpen
  \bibfield  {author} {\bibinfo {author} {\bibfnamefont {F.}~\bibnamefont
  {Ferreira}}, \bibinfo {author} {\bibfnamefont {V.~V.}\ \bibnamefont
  {Enaldiev}}, \bibinfo {author} {\bibfnamefont {V.~I.}\ \bibnamefont
  {Fal'ko}},\ and\ \bibinfo {author} {\bibfnamefont {S.~J.}\ \bibnamefont
  {Magorrian}},\ }\bibfield  {title} {\bibinfo {title} {Weak ferroelectric
  charge transfer in layer-asymmetric bilayers of 2{D} semiconductors},\ }\href
  {https://doi.org/10.1038/s41598-021-92710-1} {\bibfield  {journal} {\bibinfo
  {journal} {Scientific Reports}\ }\textbf {\bibinfo {volume} {11}},\ \bibinfo
  {pages} {13422} (\bibinfo {year} {2021}{\natexlab{a}})}\BibitemShut {NoStop}%
\bibitem [{\citenamefont {Carr}\ \emph {et~al.}(2018)\citenamefont {Carr},
  \citenamefont {Massatt}, \citenamefont {Torrisi}, \citenamefont {Cazeaux},
  \citenamefont {Luskin},\ and\ \citenamefont {Kaxiras}}]{CarrPRB2018}%
  \BibitemOpen
  \bibfield  {author} {\bibinfo {author} {\bibfnamefont {S.}~\bibnamefont
  {Carr}}, \bibinfo {author} {\bibfnamefont {D.}~\bibnamefont {Massatt}},
  \bibinfo {author} {\bibfnamefont {S.~B.}\ \bibnamefont {Torrisi}}, \bibinfo
  {author} {\bibfnamefont {P.}~\bibnamefont {Cazeaux}}, \bibinfo {author}
  {\bibfnamefont {M.}~\bibnamefont {Luskin}},\ and\ \bibinfo {author}
  {\bibfnamefont {E.}~\bibnamefont {Kaxiras}},\ }\bibfield  {title} {\bibinfo
  {title} {Relaxation and domain formation in incommensurate two-dimensional
  heterostructures},\ }\href {https://doi.org/10.1103/PhysRevB.98.224102}
  {\bibfield  {journal} {\bibinfo  {journal} {Phys. Rev. B}\ }\textbf {\bibinfo
  {volume} {98}},\ \bibinfo {pages} {224102} (\bibinfo {year}
  {2018})}\BibitemShut {NoStop}%
\bibitem [{\citenamefont {Enaldiev}\ \emph {et~al.}(2020)\citenamefont
  {Enaldiev}, \citenamefont {Z\'olyomi}, \citenamefont {Yelgel}, \citenamefont
  {Magorrian},\ and\ \citenamefont {Fal'ko}}]{Enaldiev_PRL}%
  \BibitemOpen
  \bibfield  {author} {\bibinfo {author} {\bibfnamefont {V.~V.}\ \bibnamefont
  {Enaldiev}}, \bibinfo {author} {\bibfnamefont {V.}~\bibnamefont {Z\'olyomi}},
  \bibinfo {author} {\bibfnamefont {C.}~\bibnamefont {Yelgel}}, \bibinfo
  {author} {\bibfnamefont {S.~J.}\ \bibnamefont {Magorrian}},\ and\ \bibinfo
  {author} {\bibfnamefont {V.~I.}\ \bibnamefont {Fal'ko}},\ }\bibfield  {title}
  {\bibinfo {title} {Stacking domains and dislocation networks in marginally
  twisted bilayers of transition metal dichalcogenides},\ }\href
  {https://doi.org/10.1103/PhysRevLett.124.206101} {\bibfield  {journal}
  {\bibinfo  {journal} {Phys. Rev. Lett.}\ }\textbf {\bibinfo {volume} {124}},\
  \bibinfo {pages} {206101} (\bibinfo {year} {2020})}\BibitemShut {NoStop}%
\bibitem [{\citenamefont {Enaldiev}\ \emph {et~al.}(2021)\citenamefont
  {Enaldiev}, \citenamefont {Ferreira}, \citenamefont {Magorrian},\ and\
  \citenamefont {Fal'ko}}]{Enaldiev_2021}%
  \BibitemOpen
  \bibfield  {author} {\bibinfo {author} {\bibfnamefont {V.~V.}\ \bibnamefont
  {Enaldiev}}, \bibinfo {author} {\bibfnamefont {F.}~\bibnamefont {Ferreira}},
  \bibinfo {author} {\bibfnamefont {S.~J.}\ \bibnamefont {Magorrian}},\ and\
  \bibinfo {author} {\bibfnamefont {V.~I.}\ \bibnamefont {Fal'ko}},\ }\bibfield
   {title} {\bibinfo {title} {Piezoelectric networks and ferroelectric domains
  in twistronic superlattices in {WS}$_2$/{MoS}$_2$ and {WSe}$_2$/{MoSe}$_2$
  bilayers},\ }\href {https://doi.org/10.1088/2053-1583/abdd92} {\bibfield
  {journal} {\bibinfo  {journal} {2D Materials}\ }\textbf {\bibinfo {volume}
  {8}},\ \bibinfo {pages} {025030} (\bibinfo {year} {2021})}\BibitemShut
  {NoStop}%
\bibitem [{\citenamefont {Ferreira}\ \emph
  {et~al.}(2021{\natexlab{b}})\citenamefont {Ferreira}, \citenamefont
  {Magorrian}, \citenamefont {Enaldiev}, \citenamefont {Ruiz-Tijerina},\ and\
  \citenamefont {Fal'ko}}]{Ferreira_APL}%
  \BibitemOpen
  \bibfield  {author} {\bibinfo {author} {\bibfnamefont {F.}~\bibnamefont
  {Ferreira}}, \bibinfo {author} {\bibfnamefont {S.~J.}\ \bibnamefont
  {Magorrian}}, \bibinfo {author} {\bibfnamefont {V.~V.}\ \bibnamefont
  {Enaldiev}}, \bibinfo {author} {\bibfnamefont {D.~A.}\ \bibnamefont
  {Ruiz-Tijerina}},\ and\ \bibinfo {author} {\bibfnamefont {V.~I.}\
  \bibnamefont {Fal'ko}},\ }\bibfield  {title} {\bibinfo {title} {Band energy
  landscapes in twisted homobilayers of transition metal dichalcogenides},\
  }\href {https://doi.org/10.1063/5.0048884} {\bibfield  {journal} {\bibinfo
  {journal} {Applied Physics Letters}\ }\textbf {\bibinfo {volume} {118}},\
  \bibinfo {pages} {241602} (\bibinfo {year} {2021}{\natexlab{b}})}\BibitemShut
  {NoStop}%
\bibitem [{\citenamefont {Giannozzi}\ \emph {et~al.}(2009)\citenamefont
  {Giannozzi}, \citenamefont {Baroni}, \citenamefont {Bonini}, \citenamefont
  {Calandra}, \citenamefont {Car}, \citenamefont {Cavazzoni}, \citenamefont
  {Ceresoli}, \citenamefont {Chiarotti}, \citenamefont {Cococcioni},
  \citenamefont {Dabo}, \citenamefont {Corso}, \citenamefont {de~Gironcoli},
  \citenamefont {Fabris}, \citenamefont {Fratesi}, \citenamefont {Gebauer},
  \citenamefont {Gerstmann}, \citenamefont {Gougoussis}, \citenamefont
  {Kokalj}, \citenamefont {Lazzeri}, \citenamefont {Martin-Samos},
  \citenamefont {Marzari}, \citenamefont {Mauri}, \citenamefont {Mazzarello},
  \citenamefont {Paolini}, \citenamefont {Pasquarello}, \citenamefont
  {Paulatto}, \citenamefont {Sbraccia}, \citenamefont {Scandolo}, \citenamefont
  {Sclauzero}, \citenamefont {Seitsonen}, \citenamefont {Smogunov},
  \citenamefont {Umari},\ and\ \citenamefont {Wentzcovitch}}]{QE1}%
  \BibitemOpen
  \bibfield  {author} {\bibinfo {author} {\bibfnamefont {P.}~\bibnamefont
  {Giannozzi}}, \bibinfo {author} {\bibfnamefont {S.}~\bibnamefont {Baroni}},
  \bibinfo {author} {\bibfnamefont {N.}~\bibnamefont {Bonini}}, \bibinfo
  {author} {\bibfnamefont {M.}~\bibnamefont {Calandra}}, \bibinfo {author}
  {\bibfnamefont {R.}~\bibnamefont {Car}}, \bibinfo {author} {\bibfnamefont
  {C.}~\bibnamefont {Cavazzoni}}, \bibinfo {author} {\bibfnamefont
  {D.}~\bibnamefont {Ceresoli}}, \bibinfo {author} {\bibfnamefont {G.~L.}\
  \bibnamefont {Chiarotti}}, \bibinfo {author} {\bibfnamefont {M.}~\bibnamefont
  {Cococcioni}}, \bibinfo {author} {\bibfnamefont {I.}~\bibnamefont {Dabo}},
  \bibinfo {author} {\bibfnamefont {A.~D.}\ \bibnamefont {Corso}}, \bibinfo
  {author} {\bibfnamefont {S.}~\bibnamefont {de~Gironcoli}}, \bibinfo {author}
  {\bibfnamefont {S.}~\bibnamefont {Fabris}}, \bibinfo {author} {\bibfnamefont
  {G.}~\bibnamefont {Fratesi}}, \bibinfo {author} {\bibfnamefont
  {R.}~\bibnamefont {Gebauer}}, \bibinfo {author} {\bibfnamefont
  {U.}~\bibnamefont {Gerstmann}}, \bibinfo {author} {\bibfnamefont
  {C.}~\bibnamefont {Gougoussis}}, \bibinfo {author} {\bibfnamefont
  {A.}~\bibnamefont {Kokalj}}, \bibinfo {author} {\bibfnamefont
  {M.}~\bibnamefont {Lazzeri}}, \bibinfo {author} {\bibfnamefont
  {L.}~\bibnamefont {Martin-Samos}}, \bibinfo {author} {\bibfnamefont
  {N.}~\bibnamefont {Marzari}}, \bibinfo {author} {\bibfnamefont
  {F.}~\bibnamefont {Mauri}}, \bibinfo {author} {\bibfnamefont
  {R.}~\bibnamefont {Mazzarello}}, \bibinfo {author} {\bibfnamefont
  {S.}~\bibnamefont {Paolini}}, \bibinfo {author} {\bibfnamefont
  {A.}~\bibnamefont {Pasquarello}}, \bibinfo {author} {\bibfnamefont
  {L.}~\bibnamefont {Paulatto}}, \bibinfo {author} {\bibfnamefont
  {C.}~\bibnamefont {Sbraccia}}, \bibinfo {author} {\bibfnamefont
  {S.}~\bibnamefont {Scandolo}}, \bibinfo {author} {\bibfnamefont
  {G.}~\bibnamefont {Sclauzero}}, \bibinfo {author} {\bibfnamefont {A.~P.}\
  \bibnamefont {Seitsonen}}, \bibinfo {author} {\bibfnamefont {A.}~\bibnamefont
  {Smogunov}}, \bibinfo {author} {\bibfnamefont {P.}~\bibnamefont {Umari}},\
  and\ \bibinfo {author} {\bibfnamefont {R.~M.}\ \bibnamefont {Wentzcovitch}},\
  }\bibfield  {title} {\bibinfo {title} {{QUANTUM} {ESPRESSO}: a modular and
  open-source software project for quantum simulations of materials},\ }\href
  {https://doi.org/10.1088/0953-8984/21/39/395502} {\bibfield  {journal}
  {\bibinfo  {journal} {Journal of Physics: Condensed Matter}\ }\textbf
  {\bibinfo {volume} {21}},\ \bibinfo {pages} {395502} (\bibinfo {year}
  {2009})}\BibitemShut {NoStop}%
\bibitem [{\citenamefont {Giannozzi}\ \emph {et~al.}(2017)\citenamefont
  {Giannozzi}, \citenamefont {Andreussi}, \citenamefont {Brumme}, \citenamefont
  {Bunau}, \citenamefont {Nardelli}, \citenamefont {Calandra}, \citenamefont
  {Car}, \citenamefont {Cavazzoni}, \citenamefont {Ceresoli}, \citenamefont
  {Cococcioni}, \citenamefont {Colonna}, \citenamefont {Carnimeo},
  \citenamefont {Corso}, \citenamefont {de~Gironcoli}, \citenamefont {Delugas},
  \citenamefont {DiStasio}, \citenamefont {Ferretti}, \citenamefont {Floris},
  \citenamefont {Fratesi}, \citenamefont {Fugallo}, \citenamefont {Gebauer},
  \citenamefont {Gerstmann}, \citenamefont {Giustino}, \citenamefont {Gorni},
  \citenamefont {Jia}, \citenamefont {Kawamura}, \citenamefont {Ko},
  \citenamefont {Kokalj}, \citenamefont {K\"{u}{\c{c}}\"{u}kbenli},
  \citenamefont {Lazzeri}, \citenamefont {Marsili}, \citenamefont {Marzari},
  \citenamefont {Mauri}, \citenamefont {Nguyen}, \citenamefont {Nguyen},
  \citenamefont {de-la Roza}, \citenamefont {Paulatto}, \citenamefont
  {Ponc{\'{e}}}, \citenamefont {Rocca}, \citenamefont {Sabatini}, \citenamefont
  {Santra}, \citenamefont {Schlipf}, \citenamefont {Seitsonen}, \citenamefont
  {Smogunov}, \citenamefont {Timrov}, \citenamefont {Thonhauser}, \citenamefont
  {Umari}, \citenamefont {Vast}, \citenamefont {Wu},\ and\ \citenamefont
  {Baroni}}]{QE2}%
  \BibitemOpen
  \bibfield  {author} {\bibinfo {author} {\bibfnamefont {P.}~\bibnamefont
  {Giannozzi}}, \bibinfo {author} {\bibfnamefont {O.}~\bibnamefont
  {Andreussi}}, \bibinfo {author} {\bibfnamefont {T.}~\bibnamefont {Brumme}},
  \bibinfo {author} {\bibfnamefont {O.}~\bibnamefont {Bunau}}, \bibinfo
  {author} {\bibfnamefont {M.~B.}\ \bibnamefont {Nardelli}}, \bibinfo {author}
  {\bibfnamefont {M.}~\bibnamefont {Calandra}}, \bibinfo {author}
  {\bibfnamefont {R.}~\bibnamefont {Car}}, \bibinfo {author} {\bibfnamefont
  {C.}~\bibnamefont {Cavazzoni}}, \bibinfo {author} {\bibfnamefont
  {D.}~\bibnamefont {Ceresoli}}, \bibinfo {author} {\bibfnamefont
  {M.}~\bibnamefont {Cococcioni}}, \bibinfo {author} {\bibfnamefont
  {N.}~\bibnamefont {Colonna}}, \bibinfo {author} {\bibfnamefont
  {I.}~\bibnamefont {Carnimeo}}, \bibinfo {author} {\bibfnamefont {A.~D.}\
  \bibnamefont {Corso}}, \bibinfo {author} {\bibfnamefont {S.}~\bibnamefont
  {de~Gironcoli}}, \bibinfo {author} {\bibfnamefont {P.}~\bibnamefont
  {Delugas}}, \bibinfo {author} {\bibfnamefont {R.~A.}\ \bibnamefont
  {DiStasio}}, \bibinfo {author} {\bibfnamefont {A.}~\bibnamefont {Ferretti}},
  \bibinfo {author} {\bibfnamefont {A.}~\bibnamefont {Floris}}, \bibinfo
  {author} {\bibfnamefont {G.}~\bibnamefont {Fratesi}}, \bibinfo {author}
  {\bibfnamefont {G.}~\bibnamefont {Fugallo}}, \bibinfo {author} {\bibfnamefont
  {R.}~\bibnamefont {Gebauer}}, \bibinfo {author} {\bibfnamefont
  {U.}~\bibnamefont {Gerstmann}}, \bibinfo {author} {\bibfnamefont
  {F.}~\bibnamefont {Giustino}}, \bibinfo {author} {\bibfnamefont
  {T.}~\bibnamefont {Gorni}}, \bibinfo {author} {\bibfnamefont
  {J.}~\bibnamefont {Jia}}, \bibinfo {author} {\bibfnamefont {M.}~\bibnamefont
  {Kawamura}}, \bibinfo {author} {\bibfnamefont {H.-Y.}\ \bibnamefont {Ko}},
  \bibinfo {author} {\bibfnamefont {A.}~\bibnamefont {Kokalj}}, \bibinfo
  {author} {\bibfnamefont {E.}~\bibnamefont {K\"{u}{\c{c}}\"{u}kbenli}},
  \bibinfo {author} {\bibfnamefont {M.}~\bibnamefont {Lazzeri}}, \bibinfo
  {author} {\bibfnamefont {M.}~\bibnamefont {Marsili}}, \bibinfo {author}
  {\bibfnamefont {N.}~\bibnamefont {Marzari}}, \bibinfo {author} {\bibfnamefont
  {F.}~\bibnamefont {Mauri}}, \bibinfo {author} {\bibfnamefont {N.~L.}\
  \bibnamefont {Nguyen}}, \bibinfo {author} {\bibfnamefont {H.-V.}\
  \bibnamefont {Nguyen}}, \bibinfo {author} {\bibfnamefont {A.~O.}\
  \bibnamefont {de-la Roza}}, \bibinfo {author} {\bibfnamefont
  {L.}~\bibnamefont {Paulatto}}, \bibinfo {author} {\bibfnamefont
  {S.}~\bibnamefont {Ponc{\'{e}}}}, \bibinfo {author} {\bibfnamefont
  {D.}~\bibnamefont {Rocca}}, \bibinfo {author} {\bibfnamefont
  {R.}~\bibnamefont {Sabatini}}, \bibinfo {author} {\bibfnamefont
  {B.}~\bibnamefont {Santra}}, \bibinfo {author} {\bibfnamefont
  {M.}~\bibnamefont {Schlipf}}, \bibinfo {author} {\bibfnamefont {A.~P.}\
  \bibnamefont {Seitsonen}}, \bibinfo {author} {\bibfnamefont {A.}~\bibnamefont
  {Smogunov}}, \bibinfo {author} {\bibfnamefont {I.}~\bibnamefont {Timrov}},
  \bibinfo {author} {\bibfnamefont {T.}~\bibnamefont {Thonhauser}}, \bibinfo
  {author} {\bibfnamefont {P.}~\bibnamefont {Umari}}, \bibinfo {author}
  {\bibfnamefont {N.}~\bibnamefont {Vast}}, \bibinfo {author} {\bibfnamefont
  {X.}~\bibnamefont {Wu}},\ and\ \bibinfo {author} {\bibfnamefont
  {S.}~\bibnamefont {Baroni}},\ }\bibfield  {title} {\bibinfo {title} {Advanced
  capabilities for materials modelling with quantum {ESPRESSO}},\ }\href
  {https://doi.org/10.1088/1361-648x/aa8f79} {\bibfield  {journal} {\bibinfo
  {journal} {Journal of Physics: Condensed Matter}\ }\textbf {\bibinfo {volume}
  {29}},\ \bibinfo {pages} {465901} (\bibinfo {year} {2017})}\BibitemShut
  {NoStop}%
\bibitem [{\citenamefont {Sohier}\ \emph {et~al.}(2017)\citenamefont {Sohier},
  \citenamefont {Calandra},\ and\ \citenamefont {Mauri}}]{CoulombCut}%
  \BibitemOpen
  \bibfield  {author} {\bibinfo {author} {\bibfnamefont {T.}~\bibnamefont
  {Sohier}}, \bibinfo {author} {\bibfnamefont {M.}~\bibnamefont {Calandra}},\
  and\ \bibinfo {author} {\bibfnamefont {F.}~\bibnamefont {Mauri}},\ }\bibfield
   {title} {\bibinfo {title} {Density functional perturbation theory for gated
  two-dimensional heterostructures: Theoretical developments and application to
  flexural phonons in graphene},\ }\href
  {https://doi.org/10.1103/PhysRevB.96.075448} {\bibfield  {journal} {\bibinfo
  {journal} {Phys. Rev. B}\ }\textbf {\bibinfo {volume} {96}},\ \bibinfo
  {pages} {075448} (\bibinfo {year} {2017})}\BibitemShut {NoStop}%
\bibitem [{\citenamefont {Kresse}\ and\ \citenamefont
  {Furthm\"uller}(1996)}]{VASP}%
  \BibitemOpen
  \bibfield  {author} {\bibinfo {author} {\bibfnamefont {G.}~\bibnamefont
  {Kresse}}\ and\ \bibinfo {author} {\bibfnamefont {J.}~\bibnamefont
  {Furthm\"uller}},\ }\bibfield  {title} {\bibinfo {title} {Efficient iterative
  schemes for {\it ab initio} total-energy calculations using a plane-wave
  basis set},\ }\href {https://doi.org/10.1103/PhysRevB.54.11169} {\bibfield
  {journal} {\bibinfo  {journal} {Physical Review B}\ }\textbf {\bibinfo
  {volume} {54}},\ \bibinfo {pages} {11169} (\bibinfo {year}
  {1996})}\BibitemShut {NoStop}%
\bibitem [{\citenamefont {Meyer}\ and\ \citenamefont
  {Vanderbilt}(2001)}]{Vanderbilt_PRB}%
  \BibitemOpen
  \bibfield  {author} {\bibinfo {author} {\bibfnamefont {B.}~\bibnamefont
  {Meyer}}\ and\ \bibinfo {author} {\bibfnamefont {D.}~\bibnamefont
  {Vanderbilt}},\ }\bibfield  {title} {\bibinfo {title} {Ab initio study of
  {BaTiO}$_{3}$ and {PbTiO}$_{3}$ surfaces in external electric fields},\
  }\href {https://doi.org/10.1103/PhysRevB.63.205426} {\bibfield  {journal}
  {\bibinfo  {journal} {Phys. Rev. B}\ }\textbf {\bibinfo {volume} {63}},\
  \bibinfo {pages} {205426} (\bibinfo {year} {2001})}\BibitemShut {NoStop}%
\bibitem [{Note1()}]{Note1}%
  \BibitemOpen
  \bibinfo {note} {For monolayers and 2H-bilayers we used the Coulomb
  truncation in the out-of-plane direction\cite {CoulombCut}.}\BibitemShut
  {Stop}%
\bibitem [{Note2()}]{Note2}%
  \BibitemOpen
  \bibinfo {note} {Unlike Refs. \cite {santos2013,Laturia2020}, which use
  $z$-averaged electric field for computation of dielectric permitivities of
  bilayers, we express the quadratic amendment to the total energy via
  out-of-plane displacement field, conserving across every cross-section of the
  structure. This allows us to avoid uncertainties in $\alpha _{zz}^{\protect
  \rm 3R}$, which may appear at averaging of the electric field in crystals
  with a few out-of-plane unit cells \cite {Vanderbilt_PRB}. We find that the
  polarizability, computed for a TMD monolayer, its 2H and 3R bilayer, and
  thicker 2H films, linearly scales with with the number of layers. This
  observation contradicts some earlier DFT studies of dielectric susceptibility
  of TMDs \cite {santos2013,Laturia2020} which claimed a pronounce
  layer-number-dependence, but agrees with the more recent results \cite
  {Tian2020} published by some of the authors of Refs \cite
  {santos2013}.}\BibitemShut {Stop}%
\bibitem [{\citenamefont {Slizovskiy}\ \emph {et~al.}(2019)\citenamefont
  {Slizovskiy}, \citenamefont {Garcia-Ruiz}, \citenamefont {Berdyugin},
  \citenamefont {Xin}, \citenamefont {Taniguchi}, \citenamefont {Watanabe},
  \citenamefont {Geim}, \citenamefont {Drummond},\ and\ \citenamefont
  {Fal'ko}}]{Slizovskiy2021_arxiv}%
  \BibitemOpen
  \bibfield  {author} {\bibinfo {author} {\bibfnamefont {S.}~\bibnamefont
  {Slizovskiy}}, \bibinfo {author} {\bibfnamefont {A.}~\bibnamefont
  {Garcia-Ruiz}}, \bibinfo {author} {\bibfnamefont {A.~I.}\ \bibnamefont
  {Berdyugin}}, \bibinfo {author} {\bibfnamefont {N.}~\bibnamefont {Xin}},
  \bibinfo {author} {\bibfnamefont {T.}~\bibnamefont {Taniguchi}}, \bibinfo
  {author} {\bibfnamefont {K.}~\bibnamefont {Watanabe}}, \bibinfo {author}
  {\bibfnamefont {A.~K.}\ \bibnamefont {Geim}}, \bibinfo {author}
  {\bibfnamefont {N.~D.}\ \bibnamefont {Drummond}},\ and\ \bibinfo {author}
  {\bibfnamefont {V.~I.}\ \bibnamefont {Fal'ko}},\ }\bibfield  {title}
  {\bibinfo {title} {Out-of-plane dielectric susceptibility of graphene in
  twistronic and bernal bilayers},\ }\href {https://arxiv.org/abs/1912.10067}
  {\bibfield  {journal} {\bibinfo  {journal} {arXiv:1912.10067}\ } (\bibinfo
  {year} {2019})}\BibitemShut {NoStop}%
\bibitem [{\citenamefont {Tian}\ \emph {et~al.}(2020)\citenamefont {Tian},
  \citenamefont {Scullion}, \citenamefont {Hughes}, \citenamefont {Li},
  \citenamefont {Shih}, \citenamefont {Coleman}, \citenamefont {Chhowalla},\
  and\ \citenamefont {Santos}}]{Tian2020}%
  \BibitemOpen
  \bibfield  {author} {\bibinfo {author} {\bibfnamefont {T.}~\bibnamefont
  {Tian}}, \bibinfo {author} {\bibfnamefont {D.}~\bibnamefont {Scullion}},
  \bibinfo {author} {\bibfnamefont {D.}~\bibnamefont {Hughes}}, \bibinfo
  {author} {\bibfnamefont {L.~H.}\ \bibnamefont {Li}}, \bibinfo {author}
  {\bibfnamefont {C.-J.}\ \bibnamefont {Shih}}, \bibinfo {author}
  {\bibfnamefont {J.}~\bibnamefont {Coleman}}, \bibinfo {author} {\bibfnamefont
  {M.}~\bibnamefont {Chhowalla}},\ and\ \bibinfo {author} {\bibfnamefont
  {E.~J.~G.}\ \bibnamefont {Santos}},\ }\bibfield  {title} {\bibinfo {title}
  {Electronic polarizability as the fundamental variable in the dielectric
  properties of two-dimensional materials},\ }\href
  {https://doi.org/10.1021/acs.nanolett.9b02982} {\bibfield  {journal}
  {\bibinfo  {journal} {Nano Letters}\ }\textbf {\bibinfo {volume} {20}},\
  \bibinfo {pages} {841} (\bibinfo {year} {2020})}\BibitemShut {NoStop}%
\bibitem [{\citenamefont {Slizovskiy}\ \emph {et~al.}(2021)\citenamefont
  {Slizovskiy}, \citenamefont {Garcia-Ruiz}, \citenamefont {Berdyugin},
  \citenamefont {Xin}, \citenamefont {Taniguchi}, \citenamefont {Watanabe},
  \citenamefont {Geim}, \citenamefont {Drummond},\ and\ \citenamefont
  {Fal'ko}}]{Slizovskiy2021}%
  \BibitemOpen
  \bibfield  {author} {\bibinfo {author} {\bibfnamefont {S.}~\bibnamefont
  {Slizovskiy}}, \bibinfo {author} {\bibfnamefont {A.}~\bibnamefont
  {Garcia-Ruiz}}, \bibinfo {author} {\bibfnamefont {A.~I.}\ \bibnamefont
  {Berdyugin}}, \bibinfo {author} {\bibfnamefont {N.}~\bibnamefont {Xin}},
  \bibinfo {author} {\bibfnamefont {T.}~\bibnamefont {Taniguchi}}, \bibinfo
  {author} {\bibfnamefont {K.}~\bibnamefont {Watanabe}}, \bibinfo {author}
  {\bibfnamefont {A.~K.}\ \bibnamefont {Geim}}, \bibinfo {author}
  {\bibfnamefont {N.~D.}\ \bibnamefont {Drummond}},\ and\ \bibinfo {author}
  {\bibfnamefont {V.~I.}\ \bibnamefont {Fal'ko}},\ }\bibfield  {title}
  {\bibinfo {title} {Out-of-plane dielectric susceptibility of graphene in
  twistronic and bernal bilayers},\ }\href
  {https://doi.org/10.1021/acs.nanolett.1c02211} {\bibfield  {journal}
  {\bibinfo  {journal} {Nano Letters}\ }\textbf {\bibinfo {volume} {21}},\
  \bibinfo {pages} {6678} (\bibinfo {year} {2021})}\BibitemShut {NoStop}%
\bibitem [{Note3()}]{Note3}%
  \BibitemOpen
  \bibinfo {note} {The expression for interaction energy of the FE polarization
  with displacement field naturally comes when assuming a local dielectric
  permittivity in a continuum medium approximation. Indeed, suppose the FE
  charges, with plane-averaged density $\delta \rho (z)$, are placed in the
  medium with local dielectric permittivity $\epsilon _{zz}(z)$. From the
  Poisson equation and electro-neutrality condition $\DOTSI \intop \ilimits@
  _{-\infty }^{+\infty }\delta \rho (z)dz=0$ \protect \mbox {$\partial
  _{z}(\epsilon _{zz}(z)\partial _z\varphi (z))=-\delta \rho (z)/\epsilon _0$},
  we express the potential drop across the layer of charges as $\Delta =\DOTSI
  \intop \ilimits@ _{-\infty }^{+\infty }\partial _z\varphi (z)dz=-\DOTSI
  \intop \ilimits@ _{-\infty }^{+\infty }dz\DOTSI \intop \ilimits@ _{-\infty
  }^{z}dz'\delta \rho (z')/\epsilon _{zz}(z)\epsilon _0=\DOTSI \intop \ilimits@
  _{-\infty }^{+\infty }dz\DOTSI \intop \ilimits@ ^{+\infty }_{z}dz'\delta \rho
  (z')/\epsilon _{zz}(z)\epsilon _0=\DOTSI \intop \ilimits@ \DOTSI \intop
  \ilimits@ _{z<z'}dzdz'\delta \rho (z')/\epsilon _{zz}(z)\epsilon _0$. At the
  same time, interaction energy of these charges with uniform external
  out-of-plane displacement field (related to local electric field as \protect
  \mbox {$D=\epsilon _0\epsilon _{zz}(z)E(z)$}) reads as $\delta U=-\DOTSI
  \intop \ilimits@ _{-\infty }^{+\infty }dz\delta \rho (z)\DOTSI \intop
  \ilimits@ _{-\infty }^zdz'D/\epsilon _{zz}(z')\epsilon _0= -D\DOTSI \intop
  \ilimits@ _{-\infty }^{+\infty }dz'\DOTSI \intop \ilimits@ _{-\infty
  }^{z'}dz\delta \rho (z')/\epsilon _{zz}(z)\epsilon _0=-D\DOTSI \intop
  \ilimits@ \DOTSI \intop \ilimits@ _{z<z'}dzdz'\delta \rho (z')/\epsilon
  _{zz}(z)\epsilon _0\equiv -D\Delta $. After changing variables, \protect
  \mbox {$z\leftrightarrow z'$}, at the last step of that calculation, we
  arrive at the relation in Eq. \protect \textup {\hbox {\mathsurround \z@
  \protect \normalfont (\ignorespaces \ref {Eq:interaction_intro}\unskip
  \@@italiccorr )}}.}\BibitemShut {Stop}%
\bibitem [{\citenamefont {Perdew}\ and\ \citenamefont {Zunger}(1981)}]{PBE}%
  \BibitemOpen
  \bibfield  {author} {\bibinfo {author} {\bibfnamefont {J.~P.}\ \bibnamefont
  {Perdew}}\ and\ \bibinfo {author} {\bibfnamefont {A.}~\bibnamefont
  {Zunger}},\ }\bibfield  {title} {\bibinfo {title} {Self-interaction
  correction to density-functional approximations for many-electron systems},\
  }\href {https://doi.org/10.1103/PhysRevB.23.5048} {\bibfield  {journal}
  {\bibinfo  {journal} {Phys. Rev. B}\ }\textbf {\bibinfo {volume} {23}},\
  \bibinfo {pages} {5048} (\bibinfo {year} {1981})}\BibitemShut {NoStop}%
\bibitem [{Note4()}]{Note4}%
  \BibitemOpen
  \bibinfo {note} {At small $D$ (when \protect \mbox {$y'^2\ll 1$}) approximate
  solution of Eq. \protect \textup {\hbox {\mathsurround \z@ \protect
  \normalfont (\ignorespaces \ref {Eq:differential_eq}\unskip \@@italiccorr )}}
  is given by parabola: \protect \mbox {$y(x)=2D\Delta (\ell -x)x/(\protect
  \bar {w}+2\protect \widetilde {w})$.}}\BibitemShut {Stop}%
\bibitem [{\citenamefont {Beal}\ \emph {et~al.}(2018)\citenamefont {Beal},
  \citenamefont {Hill}, \citenamefont {Martin},\ and\ \citenamefont
  {Hedengren}}]{gekko}%
  \BibitemOpen
  \bibfield  {author} {\bibinfo {author} {\bibfnamefont {L.~D.}\ \bibnamefont
  {Beal}}, \bibinfo {author} {\bibfnamefont {D.~C.}\ \bibnamefont {Hill}},
  \bibinfo {author} {\bibfnamefont {R.~A.}\ \bibnamefont {Martin}},\ and\
  \bibinfo {author} {\bibfnamefont {J.~D.}\ \bibnamefont {Hedengren}},\
  }\bibfield  {title} {\bibinfo {title} {Gekko optimization suite},\ }\href
  {https://doi.org/doi: 10.3390/pr6080106} {\bibfield  {journal} {\bibinfo
  {journal} {Processes}\ }\textbf {\bibinfo {volume} {6}},\ \bibinfo {pages}
  {106} (\bibinfo {year} {2018})}\BibitemShut {NoStop}%
\bibitem [{\citenamefont {Santos}\ and\ \citenamefont
  {Kaxiras}(2013)}]{santos2013}%
  \BibitemOpen
  \bibfield  {author} {\bibinfo {author} {\bibfnamefont {E.~J.}\ \bibnamefont
  {Santos}}\ and\ \bibinfo {author} {\bibfnamefont {E.}~\bibnamefont
  {Kaxiras}},\ }\bibfield  {title} {\bibinfo {title} {Electrically driven
  tuning of the dielectric constant in {MoS}$_2$ layers},\ }\href
  {https://doi.org/10.1021/nn403738b} {\bibfield  {journal} {\bibinfo
  {journal} {ACS Nano}\ }\textbf {\bibinfo {volume} {7}},\ \bibinfo {pages}
  {10741} (\bibinfo {year} {2013})}\BibitemShut {NoStop}%
\bibitem [{\citenamefont {Laturia}\ \emph {et~al.}(2020)\citenamefont
  {Laturia}, \citenamefont {Van~de Put},\ and\ \citenamefont
  {Vandenberghe}}]{Laturia2020}%
  \BibitemOpen
  \bibfield  {author} {\bibinfo {author} {\bibfnamefont {A.}~\bibnamefont
  {Laturia}}, \bibinfo {author} {\bibfnamefont {M.~L.}\ \bibnamefont {Van~de
  Put}},\ and\ \bibinfo {author} {\bibfnamefont {W.~G.}\ \bibnamefont
  {Vandenberghe}},\ }\bibfield  {title} {\bibinfo {title} {Dielectric
  properties of hexagonal boron nitride and transition metal dichalcogenides:
  from monolayer to bulk},\ }\href {https://doi.org/10.1038/s41699-020-00163-3}
  {\bibfield  {journal} {\bibinfo  {journal} {npj 2D Materials and
  Applications}\ }\textbf {\bibinfo {volume} {4}},\ \bibinfo {pages} {28}
  (\bibinfo {year} {2020})}\BibitemShut {NoStop}%
\end{thebibliography}%

\end{document}